\documentclass[aps,superscriptaddress,preprintnumbers,showpacs]{revtex4}
\usepackage{graphicx,wrapfig,amssymb}
\usepackage[T1]{fontenc}
\usepackage{ae,aecompl}

\newcommand{\be}{\begin{equation}}
\newcommand{\ee}{\end{equation}}
\newcommand{\ba}{\begin{eqnarray}}
\newcommand{\ea}{\end{eqnarray}}
\newcommand{\la}{\langle}
\newcommand{\ra}{\rangle}
\newcommand{\di}{ {\rm d} }

\begin{document}
\newcommand*{\PennState}{Penn State University, 104 Davey Lab, 
University Park PA 16802, U.S.A.}\affiliation{\PennState}
\newcommand*{\Dubna}{Joint Institute for Nuclear Research, Dubna, 
141980 Russia}\affiliation{\Dubna}
\newcommand*{\Bochum}{Institut f{\"u}r Theoretische Physik II, 
Ruhr-Universit{\"a}t Bochum, D-44780 Bochum, Germany}\affiliation{\Bochum}

\title{Sivers effect in semi-inclusive deeply inelastic scattering}
\author{J.~C.~Collins}\affiliation{\PennState}\affiliation{\Bochum}
\author{A.~V.~Efremov}\affiliation{\Dubna}
\author{K.~Goeke}\affiliation{\Bochum}
\author{S.~Menzel}\affiliation{\Bochum}
\author{A.~Metz}\affiliation{\Bochum}
\author{P.~Schweitzer}\affiliation{\Bochum}
\date{November 2005}


\begin{abstract}
  The Sivers function is extracted from HERMES data on single spin
  asymmetries in semi-inclusive deeply inelastic scattering.  Our
  analysis use a simple Gaussian model for the distribution of transverse
  parton momenta, together with the flavor
  dependence given by the leading $1/N_c$ approximation and a neglect
  of the Sivers antiquark distribution.  We find that within the
  errors of the data these approximations are sufficient.
\end{abstract}

\maketitle

\section{Introduction}
\label{Sec-1:introduction}

Single spin asymmetries (SSA) in hard reactions have a long history dating 
back to the 1970s when significant polarizations of $\Lambda$-hyperons in 
collisions of unpolarized hadrons were observed \cite{Bunce:1976yb}, 
and to the early 1990s when large asymmetries in 
$p^\uparrow p\to\pi X$ or $p^\uparrow \bar{p}\to\pi X$ 
were found at FNAL \cite{Adams:1991rw}. 
No fully consistent and satisfactory unifying approach to the 
theoretical description of these observations has been found so far ---
see the reviews \cite{Felix:1999tf}. 

Interestingly, the most recently observed SSA phenomena, 
namely those in semi-inclusive deeply inelastic scattering 
(SIDIS) \cite{Avakian:1999rr,Airapetian:1999tv,Airapetian:2002mf,Avakian:2003pk,HERMES-new,Airapetian:2004tw,Alexakhin:2005iw,Diefenthaler:2005gx},
seem better under control. This is in particular the case for the 
transverse target SSA observed at HERMES and COMPASS 
\cite{HERMES-new,Airapetian:2004tw,Alexakhin:2005iw,Diefenthaler:2005gx}.
On the basis of a generalized factorization approach in which transverse 
parton momenta are taken into account
\cite{Collins:1981uk,Ji:2004wu,Collins:2004nx} these ``leading twist'' 
asymmetries can be explained \cite{Boer:1997nt} in terms of the Sivers 
\cite{Sivers:1989cc,Brodsky:2002cx,Collins:2002kn,Belitsky:2002sm} or 
Collins effect \cite{Collins:1992kk}.
The former describes, loosely speaking, the distribution of unpolarized 
partons in a transversely polarized proton, the latter describes the 
fragmentation of transversely polarized partons into unpolarized hadrons.
In the transverse target SSA these effects can be distinguished by the 
different azimuthal angle distribution of the produced hadrons:
Sivers  effect $\propto\sin(\phi-\phi_S)$, while 
Collins effect $\propto\sin(\phi+\phi_S)$, where $\phi$ and $\phi_S$ denote 
respectively the azimuthal angles of the produced hadron and the target 
polarization vector with respect to the axis defined by the hard virtual 
photon \cite{Boer:1997nt}.
Both effects have been subject to intensive phenomenological studies
in hadron-hadron-collisions
\cite{Anselmino:1994tv,Anselmino:1998yz,D'Alesio:2004up,Anselmino:2004ky,Ma:2004tr}
and in SIDIS
\cite{Efremov:2001cz,DeSanctis:2000fh,Ma:2002ns,Efremov:2003tf,Anselmino:2004ht,Efremov:2004hz,Efremov:2004tp,Anselmino:2005nn,Anselmino:2005ea,Vogelsang:2005cs}.
For the longitudinal target SSA in SIDIS, which were observed first 
\cite{Avakian:1999rr,Airapetian:1999tv,Airapetian:2002mf} but are
dominated by subleading-twist effects \cite{Airapetian:2005jc}, the situation
is less clear and their description ({\sl presuming} the factorization 
theorems \cite{Collins:1981uk,Ji:2004wu,Collins:2004nx} can be generalized 
to twist-3) is more involved \cite{Mulders:1995dh,Afanasev:2003ze}.

In this work we will concentrate on the Sivers effect, which is quantified 
by the ``Sivers function'' $f_{1T}^{\perp a}(x,{\bf p}_T^2)$ (in the notation 
recommended in \cite{Bacchetta:2004jz}). It is referred to as ``naively'' or 
``artificially time-reversal-odd'' for it arises from a correlation 
between the nucleon spin ${\bf S}_{\rm T}$ and the intrinsic 
transverse parton momentum ${\bf p}_{\rm T}$, both transverse with respect 
to the nucleon momentum ${\bf P}_N$ in the infinite momentum frame,
with the effect being proportional to 
$({\bf S}_{\rm T}\times{\bf p}_{\rm T})\cdot {\bf P}_N$.

By adequately weighting the events entering the spin asymmetry with 
$\sin(\phi-\phi_S)$ one can project out from the data the Sivers SSA.
Including into the weight in addition to that a power of the transverse 
momentum of the produced hadron $P_{h\perp}\equiv |{\bf P}_{h\perp}|$ yields 
an SSA which is described 
(with the neglect of soft factors
\cite{Collins:1981uk,Ji:2004wu,Collins:2004nx}) 
model-independently in terms of the ``transverse moment'' 
$f_{1T}^{\perp (1) a}(x)$ of the Sivers function \cite{Boer:1997nt}.
{\sl Preliminary} HERMES data analyzed in this way were presented 
\cite{HERMES-new} and subject to first studies \cite{Efremov:2004tp}. 
However, the currently available {\sl final} 
HERMES and COMPASS data \cite{Airapetian:2004tw,Alexakhin:2005iw} were 
analyzed without a transverse momentum weight, and can only be interpreted 
by resorting to some {\sl model} for the distribution of the transverse 
parton momenta in the ``unintegrated'' \cite{Collins:2003fm} Sivers 
distribution and unpolarized fragmentation function. 
Different models have been explored in literature 
\cite{Anselmino:2005nn,Anselmino:2005ea,Vogelsang:2005cs}. Here 
we approximate the distribution of transverse parton
momenta in the Sivers function to be Gaussian.

We pay particular attention to the demonstration of the phenomenological
consistency of the approach, and fix or constrain 
the free parameters in the Gaussian ansatz consistently by the SIDIS HERMES 
data. Although hereby the Gaussian width of the Sivers function remains
poorly constrained, this does not prevent a meaningful 
extraction of the {\sl transverse moment} of the Sivers distribution 
function from the data \cite{Airapetian:2004tw}. This demonstrates that 
--- within the accuracy of the present data --- the Gaussian ansatz is robust 
and reliable. A comparison to extractions of the Sivers function, where no 
\cite{Efremov:2004tp} or other models 
\cite{Anselmino:2005nn,Anselmino:2005ea,Vogelsang:2005cs}
were assumed, helps to estimate the effects of model-dependence.
We find them smaller than the statistical accuracy of the present data.

In order to reduce the number of free parameters in the ansatz for the
Sivers function, we impose as an additional theoretical constraint the 
predictions for the Sivers function from the QCD limit of a large number
of colours $N_c$ \cite{Pobylitsa:2003ty}, which state that 
$f_{1T}^{\perp u}= -f_{1T}^{\perp d}$ up to $1/N_c$-corrections.
Since a fit constrained in this way works and describes the HERMES data 
\cite{Airapetian:2004tw}, our study, as a byproduct, not only tests the 
large-$N_c$ results \cite{Pobylitsa:2003ty} but also naturally explains 
the smallness of the Sivers effect from a deuteron target observed by COMPASS 
\cite{Alexakhin:2005iw}. 
Besides choosing different models for transverse parton momenta and/or
ways to fix the respective parameters, the explicit use of the
large-$N_c$ constraints is the main difference of our approach
compared to the works \cite{Anselmino:2005nn,Anselmino:2005ea,Vogelsang:2005cs}.

For the fit we use only the HERMES data \cite{Airapetian:2004tw} on the 
$x$-dependence of the Sivers SSA. Thus, the HERMES data on the $z$-dependence
serve as cross and consistency checks for the fit and the Gaussian ansatz.
We also explicitly address the questions what could be the effects of 
$1/N_c$-corrections and Sivers antiquarks (which we neglect in our fit).
Finally, we suggest how the Gaussian ansatz could be further tested by
means of SIDIS data. Such tests are of importance for they allow to
understand the range of applicability and the limitations of this ansatz.

\section{Sivers effect in SIDIS}

\begin{wrapfigure}{RD}{6cm}
        \centering
        \includegraphics[width=6cm]{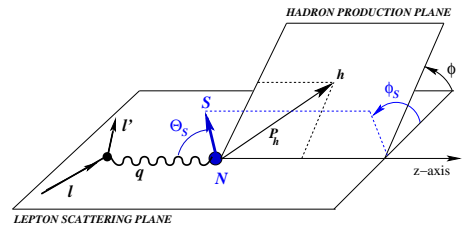}
        \caption{\label{fig1-processes-kinematics}\footnotesize\sl
        Kinematics of the SIDIS process $lp\to l^\prime h X$
        and the definitions of the azimuthal angles in the lab frame.}
        \end{wrapfigure}

Consider the process $lp^\uparrow\to l'h X$, where ``$^\uparrow$'' 
denotes the transverse (with respect to the beam) target polarization. 
Let $P$, $l$ ($l^\prime$) and $P_h$ denote respectively the momentum 
of the target proton, incoming (outgoing) lepton and produced hadron. 
The relevant kinematic variables are $q= l-l'$ with $Q^2=- q^2$, 
$x=Q^2/(2P\cdot q)$ and $z=P\cdot P_h / P\cdot q$. 
The Sivers SSA as presented in Ref.~\cite{Airapetian:2004tw} is defined
as sum over events $i$ as follows:
\be\label{Eq:AUT-Siv-unw-SIDIS-exp}
    A_{UT}^{\sin(\phi-\phi_S)}
    =
    \frac{\sum_i\sin(\phi_i-\phi_{S,i}) \,
      \left[ N^\uparrow(\phi_i;\phi_{S,i})-N^\downarrow(\phi_i;\phi_{S,i}+\pi) \right] }
    {\frac12\sum_i
     \left[ N^\uparrow(\phi_i;\phi_{S,i})+N^\downarrow(\phi_i;\phi_{S,i}+\pi) \right] } \,,
\ee
where $N^{\uparrow(\downarrow)}(\phi_i;\phi_{S,i})$ are the event counts for 
the respective target polarization (corrected for depolarization
effects) ---
see Fig.~\ref{fig1-processes-kinematics} for the definition of kinematics. 
It is understood that if the SSA is considered as function of one kinematic 
variable, then an appropriate averaging over the other variables is implied.

In order to describe the Sivers SSA as defined in 
(\ref{Eq:AUT-Siv-unw-SIDIS-exp}) we will make two major simplifications. 
The first consists in neglecting the soft factors
\cite{Collins:1981uk,Ji:2004wu,Collins:2004nx}, and their associated
energy dependence.
This step simplifies the description of the process considerably, though
it is difficult to quantify the uncertainty we introduce in this way.
Then, to leading order in the hard scale, the SSA is given by
\ba\label{Eq:AUT-Siv-unw-SIDIS-0}
    &&  A_{UT}^{\sin(\phi_h-\phi_S)}
        =
        (-2)\;\times\\
    &&  \frac{\sum_a e_a^2\,
        \int\di^2{\bf P}_{h\perp}\int\di^2{\bf p}_T\int\di^2{\bf K}_T
        \sin(\phi_h-\phi_S)\,\sin(\phi_{{\bf p}_T}-\phi_S)\,
        \frac{|{\bf p}_T|}{M_N}\,
        \delta^{(2)}(z{\bf p}_T+{\bf K}_T-{\bf P}_{h\perp})\,
        x f_{1T}^{\perp a}(x,{\bf p}_T^2)\, 
        D_1^{a}(z,{\bf K}_T^2)}{
              \sum_a e_a^2\,x f_1^a(x)\,D_1^{a}(z)} \,, \nonumber
\ea
where we cancelled out the $Q^2$ and $y$-dependent factors (which describe 
the unpolarized partonic subprocess and are the same in the numerator and 
denominator), and $\phi_{{\bf p}_T}$ denotes the azimuthal angle around 
the $z$-axis ${\bf e}_3$ of the parton struck from the target nucleon. 
In a full description of the process the transverse parton momenta in 
$f_{1T}^{\perp a}(x,{\bf p}_T^2)$ and $D_1^{a}(z,{\bf K}_T^2)$ 
would be convoluted with the soft factors 
\cite{Collins:1981uk,Ji:2004wu,Collins:2004nx} instead of the simplifying
$\delta$-function.

However, this simplification is not yet sufficient for the purpose of
extracting the Sivers function. In the numerator on the right-hand-side of 
Eq.~(\ref{Eq:AUT-Siv-unw-SIDIS-0}) the integrals convoluting the transverse 
momenta cannot be solved, unless one knows
$f_{1T}^{\perp a}(x,{\bf p}_T^2)$ and $D_1^{a}(z,{\bf K}_T^2)$,
which is not the case. The situation would be different if in the
SSA in Eq.~(\ref{Eq:AUT-Siv-unw-SIDIS-exp}) in addition had been introduced 
a power of the transverse hadron momentum $|{\bf P}_{h\perp}|$.

The leading order expression for such a weighted SSA is given by
\cite{Boer:1997nt}
\be\label{Eq:04}
        A_{UT,\pi}^{\sin(\phi-\phi_S)\frac{P_{h\perp}}{ M_N}}(x)= (-2) \;
        \frac{\sum_a e_a^2\,x f_{1T}^{\perp(1)a}(x)\,z D_1^{a/\pi}(z)}{
              \sum_a e_a^2\,x f_1^a(x)\,D_1^{a/\pi}(z)} \;,
\ee
where the transverse moment of the Sivers function is defined as
\be\label{Eq:Def-Siv-transverse-mom}
        f_{1T}^{\perp(1)a}(x) \equiv \int\!\di^2{\bf p}_T\;
        \frac{{\bf p}_T^2}{2 M_N^2}\;f_{1T}^{\perp a}(x,{\bf p}_T^2) \;.
\ee
Preliminary HERMES data on the SSA weighted in this way were presented in 
\cite{HERMES-new}.
From the point of view of minimizing the model-dependence in the analysis
\cite{Efremov:2004tp} it is preferable to consider data on an appropriately 
transverse-momentum-weighted SSA \cite{Boer:1997nt}, such as in 
Eq.~(\ref{Eq:04}).
However, a wealth of data on the SSA in SIDIS have been analyzed and 
presented without the ``appropriate transverse momentum weights'' 
\cite{Avakian:1999rr,Airapetian:1999tv,Airapetian:2002mf,Avakian:2003pk,Airapetian:2004tw,Alexakhin:2005iw,Diefenthaler:2005gx}.

Many models for the transverse momentum dependence of distribution 
and fragmentation functions were considered in literature 
\cite{Efremov:2001cz,DeSanctis:2000fh,Ma:2002ns,Efremov:2003tf,Anselmino:2004ht,Efremov:2004hz,Anselmino:2005nn,Anselmino:2005ea,Vogelsang:2005cs}.
Among the most popular models is the Gaussian ansatz, which has two important 
virtues. It describes successfully the distributions of low (with respect to 
the relevant hard scale $Q$) transverse momenta in various hard
reactions ---
see, for example, Ref.~\cite{D'Alesio:2004up}.  It also allows us to perform
analytically the integrals over transverse momentum.
In the Gaussian ansatz one assumes that the transverse momentum and $x$- or 
$z$-dependence of distribution or fragmentation functions factorize, 
and that the distributions of the transverse parton momenta are Gaussian:
\ba\label{Eq:Gauss-ansatz}
        f_1^a(x,{\bf p}_T^2) & \equiv & f_1^a(x)\;
        \frac{\exp(-{\bf p}_T^2/p^2_{\rm unp})}{\pi\; p^2_{\rm unp}}
        \;,\nonumber\\
        f_{1T}^{\perp a}(x,{\bf p}_T^2) & \equiv & f_{1T}^{\perp a}(x)\;
        \frac{\exp(-{\bf p}_T^2/p^2_{\rm Siv})}{\pi \;p^2_{\rm Siv}}
        \;,\nonumber\\
        D_1^a(z,{\bf K}_T^2) & \equiv & D_1^a(z)
        \frac{\exp(-{\bf K}_T^2/K^2_{\! D_1})}{\pi \; K^2_{\! D_1}} \;. \ea
The Gaussian widths $p^2_{\rm unp}$ and $K^2_{\! D_1}$ 
are also referred to as mean square transverse momenta of the unpolarized 
distribution and fragmentation functions, respectively, since for example
\be\label{Eq:mean-square-momentum}
        \la p_T^2\ra_{\rm unp}^{\phantom{2}} \equiv
        \frac{\int\di^2{\bf p}_T\;{\bf p}_T^2\;   f_1^a(x,{\bf p}_T^2)}
             {\int\di^2{\bf p}_T\;              f_1^a(x,{\bf p}_T^2)}
        \stackrel{\rm Gauss}{=} p_{\rm unp}^2\;.
\ee
In general, the mean square transverse 
momenta could be flavour and $x$- or $z$-dependent, a possibility that
we will disregard. 
For later convenience let us also introduce the notion of a mean parton
transverse momentum generally defined, and in the Gaussian model given 
as follows:
\be\label{Eq:mean-momentum}
        \la p_T\ra_{\rm unp} \equiv
        \frac{\int\di^2{\bf p}_T\;|{\bf p}_T|\; f_1^a(x,{\bf p}_T^2)}
             {\int\di^2{\bf p}_T\;              f_1^a(x,{\bf p}_T^2)}
        \stackrel{\rm Gauss}{=} \frac{\sqrt{\pi}}{2}\;p_{\rm unp}\;.
\ee

Under the above assumptions the expression for the Sivers SSA in 
Eqs.~(\ref{Eq:AUT-Siv-unw-SIDIS-exp},~\ref{Eq:AUT-Siv-unw-SIDIS-0}) 
is given by \cite{Efremov:2003tf}
\be\label{Eq:AUT-SIDIS-Gauss}
        A_{UT}^{\sin(\phi-\phi_S)} = (-2)\; \frac{a_{\rm Gauss}
        \sum_a e_a^2\,x f_{1T}^{\perp(1)a}(x)\, D_1^{a}(z)}{
        \sum_a e_a^2\,x f_1^a(x)\,D_1^{a}(z)} \,,
        \;\;\;\mbox{with}\;\;\;
        a_{\rm Gauss} = \frac{\sqrt{\pi}}{2}\;
        \frac{M_N}{\sqrt{p^2_{\rm Siv}+K^2_{\! D_1}/z^2}}
        \;.\ee
Here we ignore the resolution cuts applied by the experiments:
$P_{h\perp}\gtrsim 50\,{\rm MeV}$ at HERMES 
\cite{Airapetian:1999tv,Airapetian:2004tw}, and
$P_{h\perp}> 100\,{\rm MeV}$ at COMPASS \cite{Alexakhin:2005iw}.
Taking such cuts into account would yield the same expression 
(\ref{Eq:AUT-SIDIS-Gauss}), however, with an $a_{\rm Gauss}$ given in terms 
of incomplete $\Gamma$-functions. The error introduced here by neglecting 
these cuts, however, is marginal as we shall estimate below.
In the Gaussian ansatz (\ref{Eq:Gauss-ansatz}) the transverse moment 
of the Sivers function (\ref{Eq:Def-Siv-transverse-mom}) is given by
\be\label{Eq:Siv-transverse-mom-in-Gauss}
        f_{1T}^{\perp(1)a}(x)      \stackrel{\rm Gauss}{=}
        \frac{p^2_{\rm Siv}}{2 M_N^2}\;f_{1T}^{\perp a}(x) \;.
\ee
The reason why in Eq.~(\ref{Eq:AUT-SIDIS-Gauss}) we prefer to work 
with $f_{1T}^{\perp(1)a}(x)$ instead of $f_{1T}^{\perp a}(x)$ will 
become clear later. Before we start to extract the Sivers function
from the HERMES data \cite{Airapetian:2004tw}, it is necessary to fix 
or constrain the free parameters $p^2_{\rm Siv}$ and $K^2_{\! D_1}$,
preferably by (other) HERMES data for sake of consistency.
The next Section is devoted to this task.

\section{\boldmath
        Unpolarized SIDIS, 
        positivity constraints, 
        and the large $N_c$-limit}
\label{Sec-3:unpol-SIDIS+positivity}

        \begin{wrapfigure}{RD!}{6cm}
        \vspace{-0.3cm}
        \centering
        \includegraphics[width=2.0in]{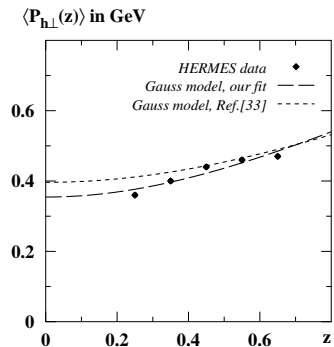}
        \caption{\footnotesize\sl
        \label{Fig2-Phperp-av}
        The average transverse momentum $\la P_{h\perp}(z)\ra$ of pions 
        produced in SIDIS as measured by HERMES from a deuterium target 
        \cite{Airapetian:2002mf} vs.\  $z$. 
        The dashed curve is the $\la P_{h\perp}(z)\ra$ in the
        Gaussian model (\ref{Eq:Phperp-av}) with the parameters as fixed
        here, see Eq.~(\ref{Eq:fit-pT2-KT2}). The dotted curve is the 
        $\la P_{h\perp}(z)\ra$ Gaussian model with the parameters as 
        obtained from a study of data on the Cahn effect 
        \cite{Anselmino:2005nn}.}
        \end{wrapfigure}

Let us assume the distribution of transverse momenta in $f_1^a(x,{\bf p}_T^2)$
and $D_1^a(z,{\bf K}_T^2)$ to be Gaussian according to (\ref{Eq:Gauss-ansatz}),
and let us furthermore assume the corresponding Gaussian widths to be flavour
and $x$- or $z$-independent. Then the average transverse momentum of the 
produced hadrons as function of $z$ is given by 
\ba\label{Eq:Phperp-av}
        \la P_{h\perp}(z)\ra 
        &=& \frac{\sqrt{\pi}}{2}\,\sqrt{z^2 p^2_{\rm unp}+K^2_{\!D_1}}\;.
\ea
Fig.~\ref{Fig2-Phperp-av} shows the HERMES data on $\la P_{h\perp}(z)\ra$
for $h=$ pions from Ref.~\cite{Airapetian:2002mf}. Strictly speaking these 
data were taken from a deuterium target, but we will ignore
this fact (i.e.,  we neglect nuclear binding effects and use isospin symmetry).
Also we will ignore the fact that these mean values are not corrected
for acceptance effects. 
What is important for us is that these data allow to fix the free parameters
$p^2_{\rm unp}$ and $K^2_{\!D_1}$. A best fit yields
\ba\label{Eq:fit-pT2-KT2}
        p^2_{\rm unp} &=& 0.33\,{\rm GeV}^2 \;,\nonumber\\
        K^2_{\! D_1}  &=& 0.16\,{\rm GeV}^2 \;,\ea
and is shown in Fig.~(\ref{Fig2-Phperp-av}) by a dashed line. 
We observe a good and for our purposes sufficient agreement. It is important 
to stress that the agreement could be improved at the prize of introducing a 
$z$- and/or flavour dependent Gaussian width for the unpolarized fragmentation
function, but we will refrain from doing so and stick to our simple picture.

It is instructive to compare our result (\ref{Eq:fit-pT2-KT2}) to the
values extracted in Ref.~\cite{Anselmino:2005nn} under certain assumptions
from EMC data \cite{Arneodo:1986cf} on the so-called Cahn effect 
\cite{Cahn:1978se}.
There, $p^2_{\rm unp}=0.25\,{\rm GeV}^2$ and $K_{\! D_1}^2=0.20\,{\rm GeV}^2$
were found. With these numbers one obtains a $\la P_{h\perp}(z)\ra$ which 
describes the HERMES data almost as well as the direct fit in
Eq.~(\ref{Eq:fit-pT2-KT2}) --- Fig.~(\ref{Fig2-Phperp-av}).
(Notice that in the formalism of Ref.~\cite{Anselmino:2005nn} the expression
(\ref{Eq:Phperp-av}) holds approximately upon the neglect of terms of 
${\cal O}(k_\perp^2/Q^2)$.)
Thus, we are lead to the encouraging conclusion that the Gaussian ansatz for 
$f_1^a(x,{\bf p}_T^2)$ and $D_1^a(z,{\bf K}_T^2)$ is compatible with SIDIS 
data from HERMES \cite{Airapetian:2002mf} and that the ansatz and the numerical
values for the Gaussian widths are supported qualitatively by the analysis
\cite{Anselmino:2005nn} of EMC data \cite{Arneodo:1986cf} on the Cahn effect.

Unlike in the case of the unpolarized fragmentation and distribution functions,
it is not possible to fix the Gaussian width of the Sivers function from 
SIDIS data. As the HERMES data \cite{Airapetian:2004tw} show a non-zero Sivers 
effect, of course, the parameter $p^2_{\rm Siv}$ cannot be zero \footnote{
        Otherwise, for $p^2_{\rm Siv}\to 0$ in (\ref{Eq:Gauss-ansatz}) the 
        Gaussian 
        $\frac{\exp(-{\bf p}_T^2/p_{\rm Siv}^2)}{\pi p_{\rm Siv}^2}\to$
        $\delta^{(2)}({\bf p}_T)$.
        Then $f_{1T}^{\perp(1)a}(x)\to 0$ and the Sivers SSA 
        (\ref{Eq:AUT-SIDIS-Gauss}) would vanish.}.
However, there is also a non-trivial {\sl upper} bound for $p_{\rm Siv}^2$ 
due to positivity conditions. The positivity bound for the Sivers distribution 
function reads \cite{Bacchetta:1999kz} 
\be\label{Eq:positivity-bound-1}
        \frac{|{\bf p}_T|}{M_N}\,|f_{1T}^{\perp a}(x,{\bf p}_T^2)| 
        \leq f_1^a(x,{\bf p}_T^2) \;.
\ee
If we demand the inequality (\ref{Eq:positivity-bound-1}) to be satisfied
in the Gaussian model (\ref{Eq:Gauss-ansatz})
at any value of $x$ and for all $|{\bf p}_T|$,
then the following necessary and sufficient condition must hold
\be\label{Eq:positivity-bound-2}
	\left(\frac{f_{1T}^{\perp a}(x)}{f_1^a(x)}\right)^{\!\!2}
	\le \frac{2e\,M_N^2}{p^2_{\rm unp}} \,R(1-R)
        \;\;\;\;\;\mbox{where}\;\;\;\;\;
	R \equiv \frac{p^2_{\rm siv}}{p^2_{\rm unp}}\,.
\ee
This means that the Gaussian width is bound from below and above as follows
\be	\frac12-\sqrt{\frac14-\frac{p^2_{\rm unp}}{2e\,M_N^2}
	\left(\frac{f_{1T}^{\perp a}(x)}{f_1^a(x)}\right)^{\!\!2}\,}
        \leq R \leq
	\frac12+\sqrt{\frac14-\frac{p^2_{\rm unp}}{2e\,M_N^2}
	\left(\frac{f_{1T}^{\perp a}(x)}{f_1^a(x)}\right)^{\!\!2}}\;.
\ee
In particular, if $f_{1T}^{\perp a}(x)\neq0$ as the data tell us, then $0<R<1$ 
must be satisfied. Thus, we see that $p^2_{\rm Siv}$ is restricted to the range
\be\label{Eq:positivity-bound-2a}
        0 < p^2_{\rm Siv}<0.33 \,{\rm GeV}^2\;.
\ee 
It is worthwhile stressing that there is a bound for $p^2_{\rm Siv}$ even if 
$f_{1T}^{\perp a}(x)$ is very small.

For later convenience let us derive from (\ref{Eq:positivity-bound-1}) 
bounds for the transverse moment of the Sivers function. 
From the Gaussian version (\ref{Eq:positivity-bound-2}) of the positivity 
bound we obtain immediately
\be\label{Eq:positivity-bound-2use}
	\left(\frac{f_{1T}^{\perp(1)a}(x)}{f_1^a(x)}\right)^{\!\!2}
	\le \frac{e\,p^2_{\rm unp}}{2M_N^2} \,R^3(1-R)
	\le \frac{e\,p^2_{\rm unp}}{2M_N^2} \,\frac{3^3}{4^4}\;.
\ee
Notice that from (\ref{Eq:positivity-bound-1}) one also can derive a
model-independent bound as follows. Using
$$
        \left| \int\di^2{\bf p}_T\;\frac{{\bf p}_T^2}{2M_N^2} 
        f_{1T}^{\perp a}(x,{\bf p}_T^2) \right| \leq 
        \int\di^2{\bf p}_T\;\frac{{\bf p}_T^2}{2M_N^2} 
        \left| f_{1T}^{\perp a}(x,{\bf p}_T^2) \right| \leq 
        \int\di^2{\bf p}_T\;\frac{|{\bf p}_T|}{2M_N} 
        f_1^a(x,{\bf p}_T^2)    
$$
and the definition (\ref{Eq:mean-momentum}) we obtain 
\be\label{Eq:positivity-bound-3}
        |f_{1T}^{\perp(1)a}(x)|\leq\frac{\la p_T\ra_{\rm unp}}{2M_N}\;f_1^a(x)
        \;. \ee
In the derivation of this bound no use was made of any transverse momentum model.
Therefore, it must be valid in any model. In fact, by evaluating $\la p_T\ra_{\rm unp}$
in the Gaussian model (\ref{Eq:mean-momentum}), we find that the model bound
(\ref{Eq:positivity-bound-2use}) is stronger than (\ref{Eq:positivity-bound-3}),
i.e.\  if the transverse moment of the Sivers functions satisfies 
(\ref{Eq:positivity-bound-2use}) then it fullfils automatically also
(\ref{Eq:positivity-bound-3}).

When using the inequality (\ref{Eq:positivity-bound-2})
in our approach with the Gaussian widths assumed to be $x$-independent, 
it is understood that the ratio of the Sivers function (or its moment) 
to $f_1^a(x)$ is to be evaluated at that point in $x$, where it takes its 
maximal value.

The present SIDIS data with their sizeable error bars 
\cite{Airapetian:2004tw,Alexakhin:2005iw} do not constrain fits for 
$f_{1T}^{\perp a}$ for the separate flavours $a=u$, $d$, $\bar{u}$ and 
$\bar{d}$ assuming the effects of heavier quarks to be negligible.
In fact, in Ref.~\cite{Anselmino:2005nn} where this has been attempted all 
fitted distributions but $f_{1T}^{\perp u}$ were found consistent with zero.
In this situation it appears appealing to invoke additional theoretical 
constraints. In particular, here we will use predictions from the QCD 
limit of a large number of colours $N_c$. 

In this limit the nucleon appears as $N_c$ quarks bound by a mean field
\cite{Witten:1979kh}, which exhibits certain spin-flavour symmetries 
\cite{Balachandran:1982cb}. By exploring these symmetry properties it 
was proven in a model independent way that in the large-$N_c$ limit
\cite{Pobylitsa:2003ty} 
\be\label{Eq:large-Nc-0}
        \underbrace{|(f_{1T}^{\perp u}-f_{1T}^{\perp d})(x,{\bf p}_T^2)|}
                  _{\displaystyle \hspace{1.2cm} = {\cal O}(N_c^3)} 
        \gg
        \underbrace{|(f_{1T}^{\perp u}+f_{1T}^{\perp d})(x,{\bf p}_T^2)|}
                  _{\displaystyle \hspace{1.2cm} = {\cal O}(N_c^2)} \;,\ee
or, equivalently,
\be\label{Eq:large-Nc}
      f_{1T}^{\perp u}(x,{\bf p}_T^2) =
    - f_{1T}^{\perp d}(x,{\bf p}_T^2) \;\;\;
    \mbox{modulo $1/N_c$ corrections.}\ee
The relations (\ref{Eq:large-Nc-0},~\ref{Eq:large-Nc}) are expected
to be valid within their accuracy in the region of not too small
and not too large $x$ satisfying $xN_c={\cal O}(N_c^0)$.
Similar relations hold for the Sivers antiquark distributions\footnote{
        For historical correctness we mention that previously 
        (\ref{Eq:large-Nc}) was discussed in the framework of (simple 
        versions of) chiral models \cite{Anselmino:2001vn}. However, 
        the way in which (\ref{Eq:large-Nc}) was obtained there was 
        shown to be incorrect \cite{Pobylitsa:2002fr}.
        Recently, in Ref.~\cite{Drago:2005gz} a (more sophisticated
        version of a) chiral model with vector mesons obeying a hidden 
        local flavour symmetry was discussed, in which the Sivers 
        function obeys (\ref{Eq:large-Nc}).}.

Inspired by the large-$N_c$ relation (\ref{Eq:large-Nc}) we choose
the following ansatz:
\be\label{Eq:ansatz}
        x\,f_{1T}^{\perp (1) u}(x) = -x\,f_{1T}^{\perp (1) d}(x)
        = A \, x^b\,(1-x)^5 \,,
\ee
and set $f_{1T}^{\perp(1)\bar q}(x)$, as well as the Sivers 
distributions of heavier quarks, to zero. The shape of the Sivers 
function at large $x$ cannot be constrained by the data 
\cite{Airapetian:2004tw,Alexakhin:2005iw}. The large-$x$ behaviour 
$f_{1T}^{\perp (1) q}(x) \propto (1-x)^5$ can be justified under
certain assumptions --- see \cite{Efremov:2004tp} and references therein.
However, one may consider it here as a mere model ansatz.

Given the size of the error bars of the present data the ansatz 
(\ref{Eq:ansatz}) and the above assumptions are not too restrictive.
This was exemplified in Ref.~\cite{Efremov:2004tp} in a study of
{\sl preliminary} HERMES data \cite{HERMES-new} on the
transverse-momentum-weighted Sivers SSA (\ref{Eq:04}).

\section{Extraction of the Sivers function 
          from HERMES data}
\label{Sec-5:Extraction}

In the previous Sections we fixed the parameter $K^2_{\! D_1}$
in Eq.~(\ref{Eq:fit-pT2-KT2}), but all we have been able to do 
so far concerning $p^2_{\rm Siv}$ was to constrain this parameter to the 
vague range in Eq.~(\ref{Eq:positivity-bound-2a}). 
However, the parameter $p^2_{\rm Siv}$ appears {\sl explicitly} 
only in the ``Gaussian factor'' $a_{\rm Gauss}$ defined in 
Eq.~(\ref{Eq:AUT-SIDIS-Gauss}).
For illustrative purposes we evaluate this factor for the 
$\la z^2\ra\approx (0.4)^2$ of the HERMES experiment in the range 
(\ref{Eq:positivity-bound-2a}) of the a priori possible value for 
$p^2_{\rm Siv}$, and we observe that it is strongly constrained to 
be in the range
\be
        0.72 < a_{\rm Gauss} < 0.83 \;.
\ee
Thus, as long as one is interested in extracting from the HERMES data 
\cite{Airapetian:2004tw} the transverse moment $f_{1T}^{\perp(1)a}(x)$ 
under the assumption of the Gaussian ansatz (\ref{Eq:Gauss-ansatz}), 
then the result is only affected to at most $\pm 10\%$ by variations of
the parameter $p^2_{\rm Siv}$ within the poorly constrained range
given in (\ref{Eq:positivity-bound-2a}).

%
\begin{figure}
\begin{tabular}{cc}
\includegraphics[width=2.2in]{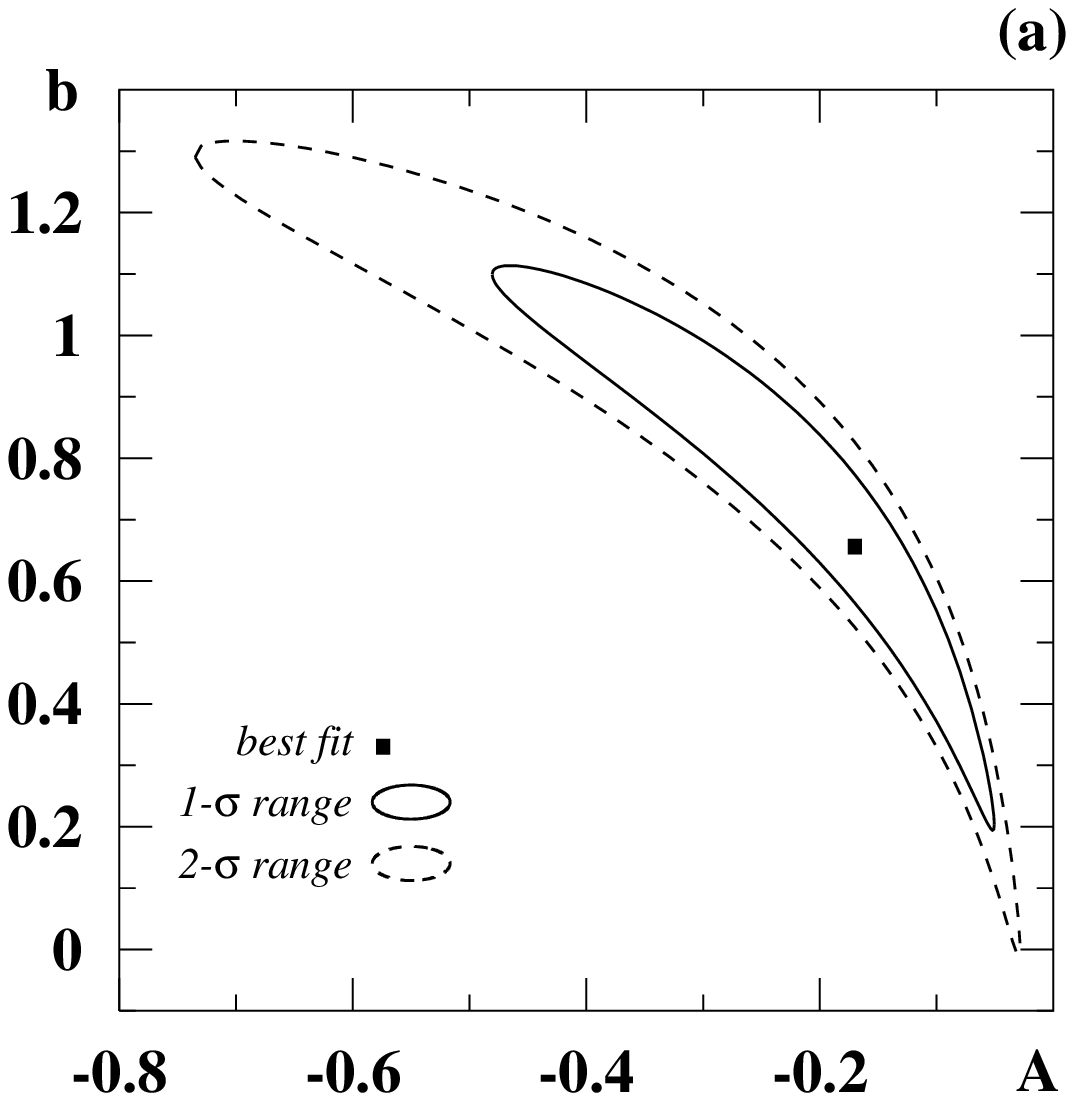}&
\includegraphics[width=2.25in]{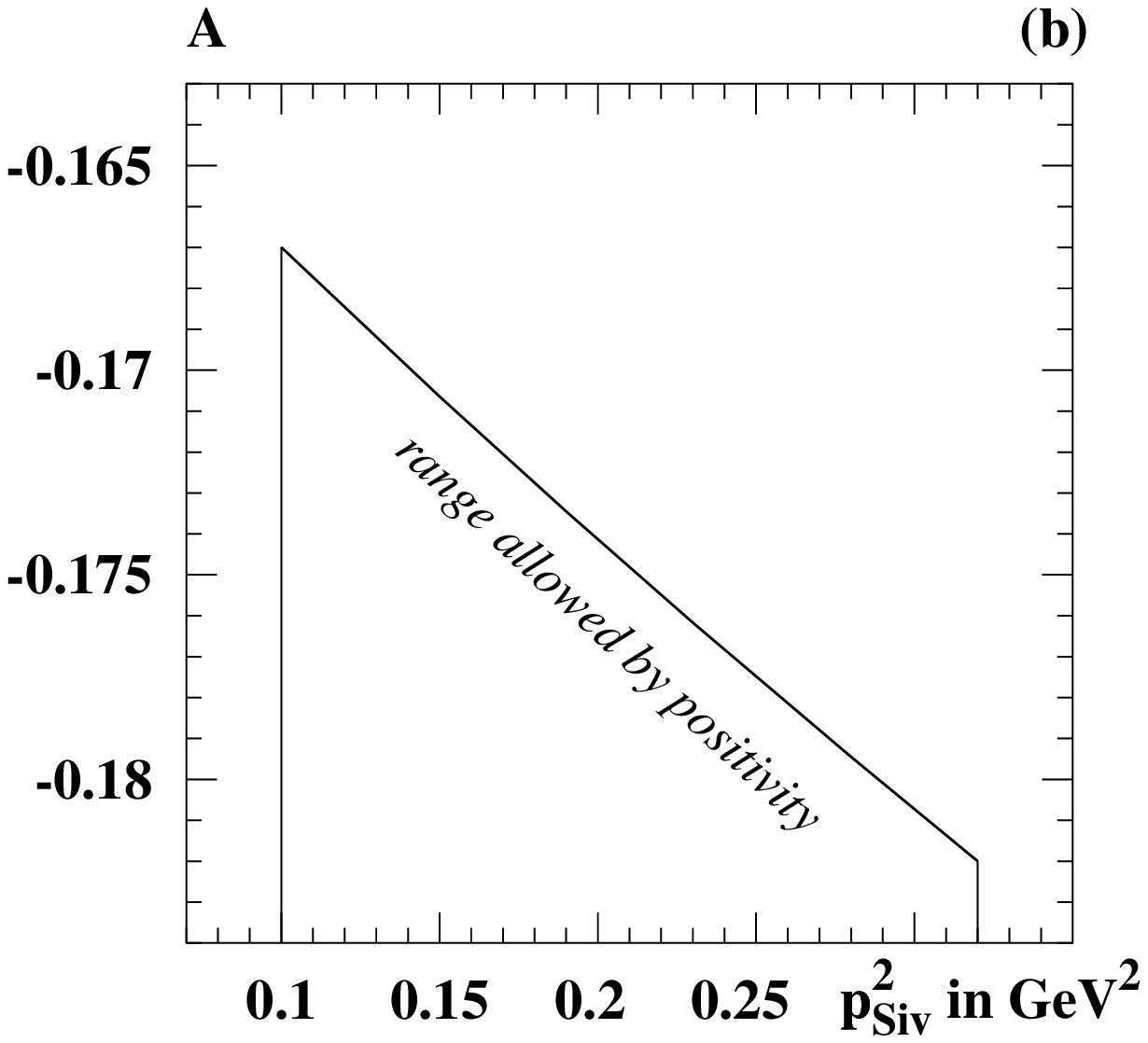}
\end{tabular}
\caption{\label{Fig3-A-B-pSiv}\footnotesize\sl
        {\bf a.}
        The best fit and the respectively 1- and 2-$\sigma$ range for the 
        parameters $A$ and $b$ in the ansatz (\ref{Eq:ansatz}) for the 
        Sivers function. 
        {\bf b.}
        The dependence of the parameter $A$ on the Gaussian width 
        $p_{\rm Siv}^2$ characterizing the transverse momentum distribution 
        in the Sivers function in the Gaussian model (\ref{Eq:Gauss-ansatz}).
        The parameter $b$ is practically $p_{\rm Siv}^2$-independent.}
\end{figure}

\begin{figure}
%
%
\begin{tabular}{cc}
\includegraphics[width=2.2in]{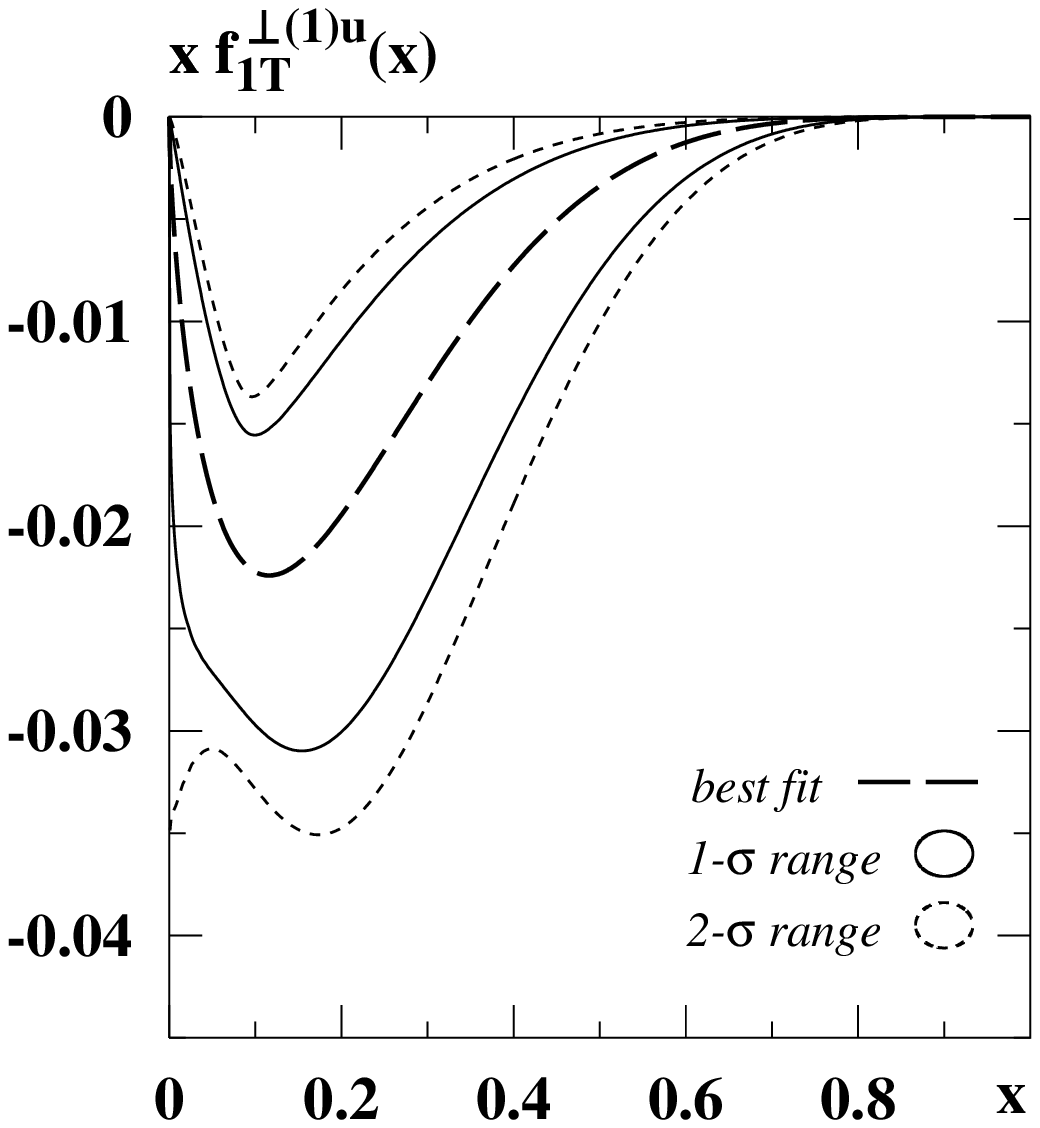}&
\includegraphics[width=2.2in]{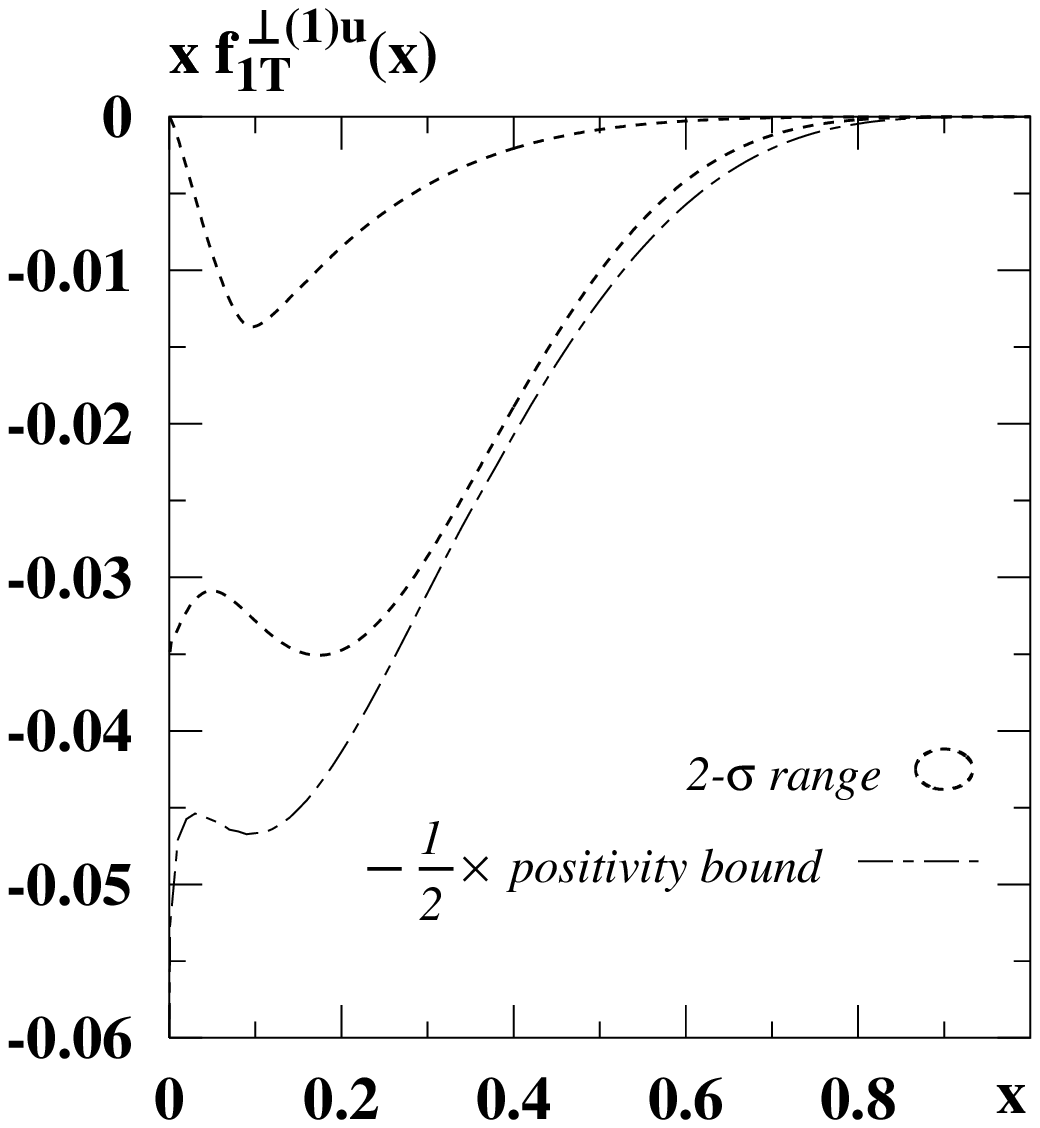}
\end{tabular}
\caption{\label{Fig4-f1Tperp-fit}\footnotesize\sl
        {\bf a.}
        The $u$-quark Sivers function $xf_{1T}^{\perp (1)u}(x)$ as function
        of $x$, as extracted from the HERMES data \cite{Airapetian:2004tw}. 
        Shown are the best fit, and its 1- and 2-$\sigma$ regions.
        {\bf b.}
        Here it is shown that the absolute value of the extracted
        Sivers function  
        does not exceed half of the positivity bound in the Gaussian model 
        in Eq.~(\ref{Eq:positivity-bound-2use}).}
\end{figure}

\begin{figure}
%
%
\begin{tabular}{cc}
\includegraphics[width=2.2in]{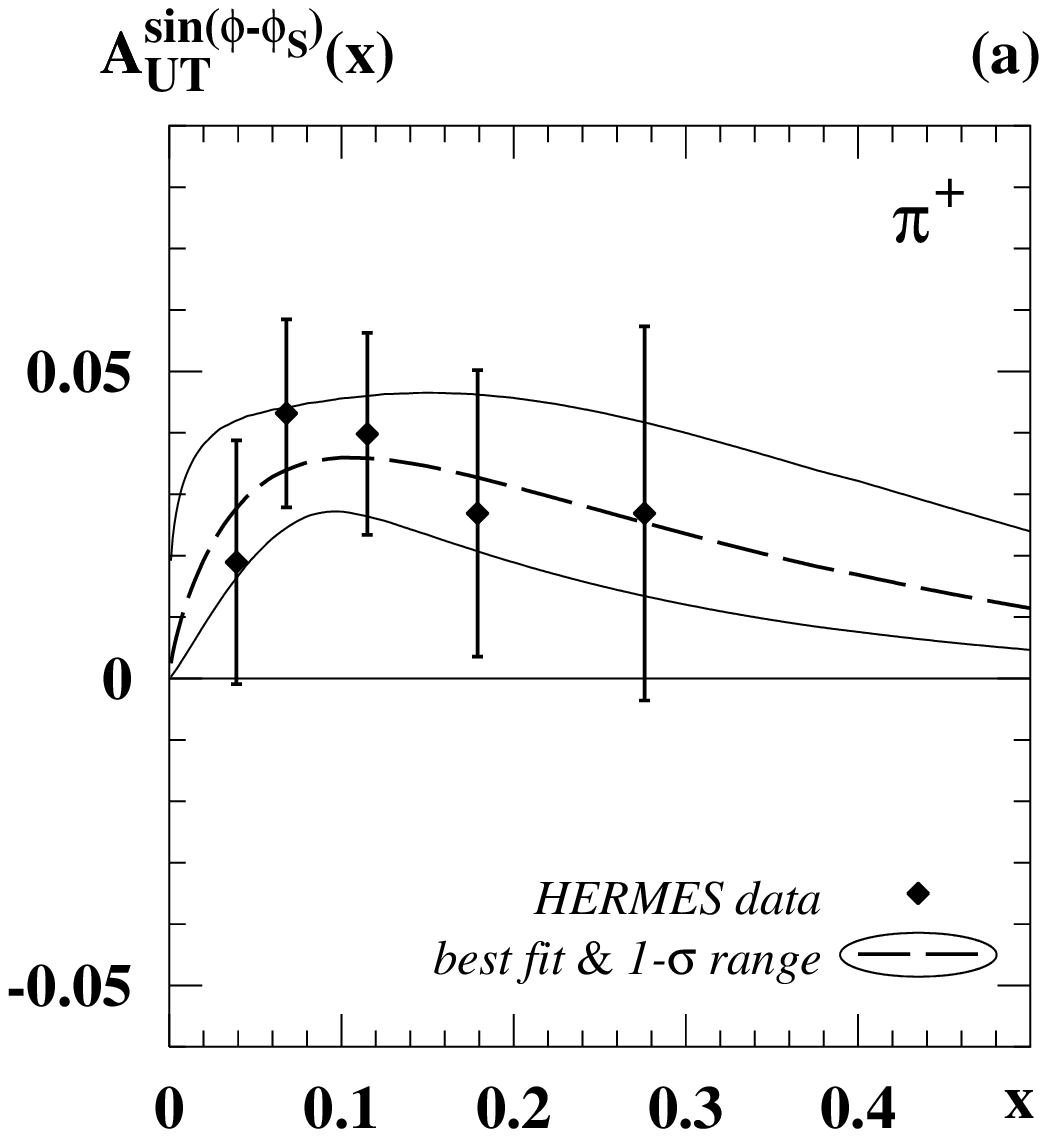}&
\includegraphics[width=2.2in]{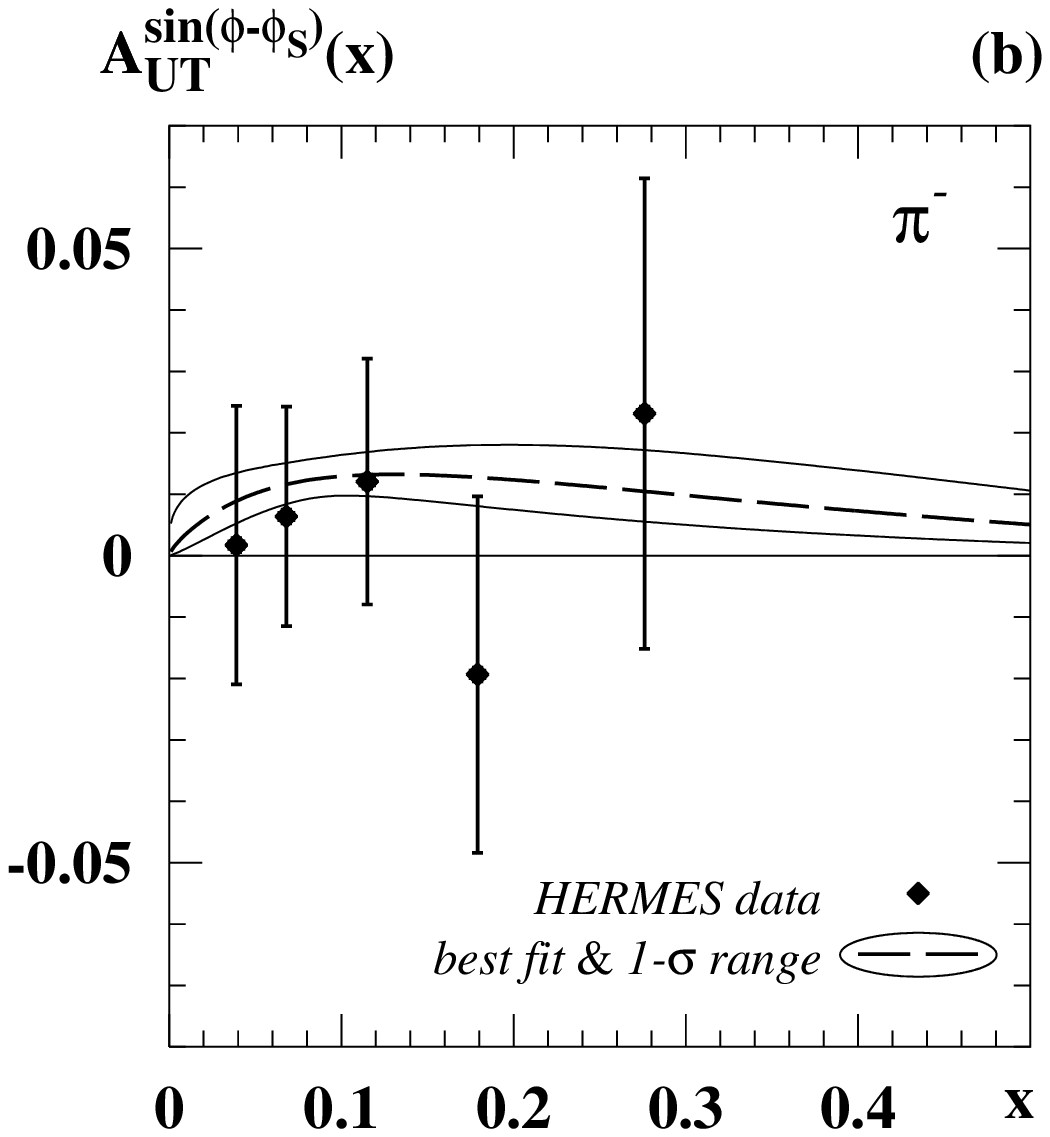}
\end{tabular}
\caption{\label{Fig5-compare-to-data}\footnotesize\sl
        The azimuthal SSA $A_{UT}^{\sin(\phi_h-\phi_S)}$ as function of $x$
        for charged pions as obtained from the fit (\ref{Eq:2nd-fit}) in 
        comparison to the {\sl final} HERMES data \cite{Airapetian:2004tw}.}
\end{figure}
%

Note that as soon as we extract a result for $f_{1T}^{\perp(1)a}(x)$, then 
from the inequality (\ref{Eq:positivity-bound-2use})
we immediately obtain bounds for  $p^2_{\rm Siv}$ which are stronger 
than those given in (\ref{Eq:positivity-bound-2a}). We proceed as follows.

We use the parameterizations \cite{Gluck:1998xa,Kretzer:2001pz} for $f_1^a(x)$
and $D_1^a(z)$ at a scale of $2.5\,{\rm GeV}^2$ which corresponds to the 
average $Q^2$ in the HERMES experiment. We insert the large-$N_c$ motivated 
ansatz (\ref{Eq:ansatz}) in the expression for the $x$-dependent Sivers SSA 
in Eq.~(\ref{Eq:AUT-SIDIS-Gauss})
\be\label{Eq:AUT-SIDIS-Gauss-extract}
        A_{UT}^{\sin(\phi-\phi_S)} = (-2)\; \frac{
        \int\di z\;a_{\rm Gauss}(z)\, 
        \sum_a e_a^2\,x f_{1T}^{\perp(1)a}(x)\, D_1^{a}(z)}{
        \int\di z\;\sum_a e_a^2\,x f_1^a(x)\,D_1^{a}(z)}
\ee
where we consider the $z$-dependence of the Gauss factor 
$a_{\rm Gauss}$ and integrate within the cuts \cite{Airapetian:2004tw} 
of the HERMES experiment, $0.2\leq z \leq 0.7$. 
We fix $K^2_{\! D_1}$ according to Eq.~(\ref{Eq:fit-pT2-KT2}) and 
choose for the parameter $p^2_{\rm Siv}$ a value out of the range 
(\ref{Eq:positivity-bound-2a}) 
of the {\sl a priori} possible values.
Then we check whether for the chosen $p^2_{\rm Siv}$ the extracted
$f_{1T}^{\perp(1)a}(x)$ does satisfy the inequality 
(\ref{Eq:positivity-bound-2use}) within its 2-$\sigma$-uncertainty.
We obtain the following result which refers to a scale of about 
$2.5\,{\rm GeV}^2$:
\ba\label{Eq:2nd-fit}
        x\,f_{1T}^{\perp(1)u}(x) = -x\,f_{1T}^{\perp(1)d}(x)
        &=& -(0.17\dots 0.18)\,x^{0.66}(1-x)^5 \;,\\
   \label{Eq:pSiv2-range}       
        \mbox{with}\;\;\;\;
        p_{\rm Siv}^2 
        &=& \phantom{-} (0.10\dots 0.32)\,{\rm GeV}^2\;.
\ea
Several comments are in order. The total $\chi^2$ is about $2.2$, i.e.,  
the $\chi^2$ per degree of freedom is about $0.3$.

In Fig.~\ref{Fig3-A-B-pSiv}a the 1- and 2-$\sigma$ regions of the parameters  
$A$ and $b$ are shown for $p_{\rm Siv}^2= 0.20 \,{\rm GeV}^2$, which 
is the central value of  $p_{\rm Siv}^2$ in the range (\ref{Eq:pSiv2-range}).
The dependence of the parameter $b$ on $p_{\rm Siv}^2$ is negligible.
The response of the parameter $A$ to variations of $p_{\rm Siv}^2$ 
is nearly linear --- Fig.~\ref{Fig3-A-B-pSiv}b.

The range (\ref{Eq:pSiv2-range}) consists of those values of $p_{\rm Siv}^2$
that are allowed by the general positivity inequality in the Gaussian model
in Eq.~(\ref{Eq:positivity-bound-2}). 
The data \cite{Airapetian:2004tw} do not allow us to constrain the parameter 
$p_{\rm Siv}^2$ more accurately than that. However, it
is satisfactory to observe how little the fit 
result for $f_{1T}^{\perp(1)a}$ in Eq.~(\ref{Eq:2nd-fit}) is affected by 
the fact that the parameter $p_{\rm Siv}^2$ is only poorly constrained 
in the range (\ref{Eq:pSiv2-range}).

In Fig.~\ref{Fig4-f1Tperp-fit}a we show the extracted $u$-quark Sivers
function 
$xf_{1T}^{\perp (1)u}(x)$. For the curve labeled as ``best fit'' we have chosen 
$p^2_{\rm Siv}= 0.20 \,{\rm GeV}^2$ which corresponds to the central 
value in the range (\ref{Eq:pSiv2-range}) allowed by positivity requirements.
For the minima (maxima) of the 1- and 2-$\sigma$ regions we have chosen the 
minimal (maximal) value of $p^2_{\rm Siv}$ in the range (\ref{Eq:pSiv2-range}).
Thus, the displayed error bands contain both, the statistical error of the 
HERMES data \cite{Airapetian:2004tw} and the uncertainty due to the poorly 
constrained Gaussian width $p^2_{\rm Siv}$ of the Sivers function in 
Eq.~(\ref{Eq:pSiv2-range}).

Notice that strictly speaking we neglected a low-$P_{h\perp}$ cut in 
Eq.~(\ref{Eq:AUT-SIDIS-Gauss}) and one may wonder how large is the error 
we introduced in this way. With the results we obtained (neglecting such cuts) 
in Eqs.~(\ref{Eq:fit-pT2-KT2},~\ref{Eq:pSiv2-range}) we find that taking 
this cut into account in the HERMES kinematics would change our results by 
about $1\%$ (and about $2\%$ at COMPASS to be discussed below).
Thus, in the Gaussian model the neglect of the low-$P_{h\perp}$ cut is justified.

The absolute values of the extracted Sivers functions
$f_{1T}^{\perp (1)u}(x) = - xf_{1T}^{\perp (1)d}(x)$ are restricted
by the upper (Gaussian model) bound (\ref{Eq:positivity-bound-2use}) given 
numerically by $0.23\,f_1^u(x)$ or $0.23\,f_1^d(x)$. 
Since $f_1^d(x)$ is smaller than $f_1^u(x)$ the bound is stronger
for the $d$-quark. In Fig.~\ref{Fig4-f1Tperp-fit}b we see 
that the extracted Sivers function well satisfies this bound,
and we remark that it does not even exceed half of the general bound 
(\ref{Eq:positivity-bound-3}) within its 2-$\sigma$ uncertainty.

Finally, in Fig.~\ref{Fig5-compare-to-data}
we compare the Sivers SSA obtained on the basis of our fit (\ref{Eq:2nd-fit}) 
to the HERMES data \cite{Airapetian:2004tw}.
Of course, in the SSA the effects of the (small) uncertainty of the parameter
$A$ in (\ref{Eq:2nd-fit}) and the (sizeable) uncertainty of the parameter 
$p^2_{\rm Siv}$ in (\ref{Eq:pSiv2-range}) cancel.
Notably, the 1-$\sigma$ error band for the $\pi^-$ SSA is much 
narrower than for the $\pi^+$ SSA. This means that the $\pi^-$ SSA is more
sensitive to $1/N_c$ corrections, i.e., to deviations from the ansatz
(\ref{Eq:ansatz}). We will discuss this point in detail in the next
Section.

As an intermediate summary we conclude that the HERMES data 
\cite{Airapetian:2004tw} are well compatible with the large-$N_c$ predictions 
(\ref{Eq:large-Nc}) for the Sivers function \cite{Pobylitsa:2003ty} and that
the fit (\ref{Eq:2nd-fit}) satisfies the positivity bounds
\cite{Bacchetta:1999kz}. 
Remarkably, the sign of the extracted Sivers function in Eq.~(\ref{Eq:2nd-fit})
is in agreement with the physical picture discussed in \cite{Burkardt:2002ks}.
We remark, however, that model calculations of the Sivers function
\cite{Bacchetta:2003rz,Lu:2004au,Yuan:2003wk} show no tendency to 
exhibit the large-$N_c$ pattern (\ref{Eq:large-Nc}).

\section{\boldmath Cross checks: HERMES data on $z$-dependence \& COMPASS data}

\begin{figure}
\begin{tabular}{cc}
\includegraphics[width=2.2in]{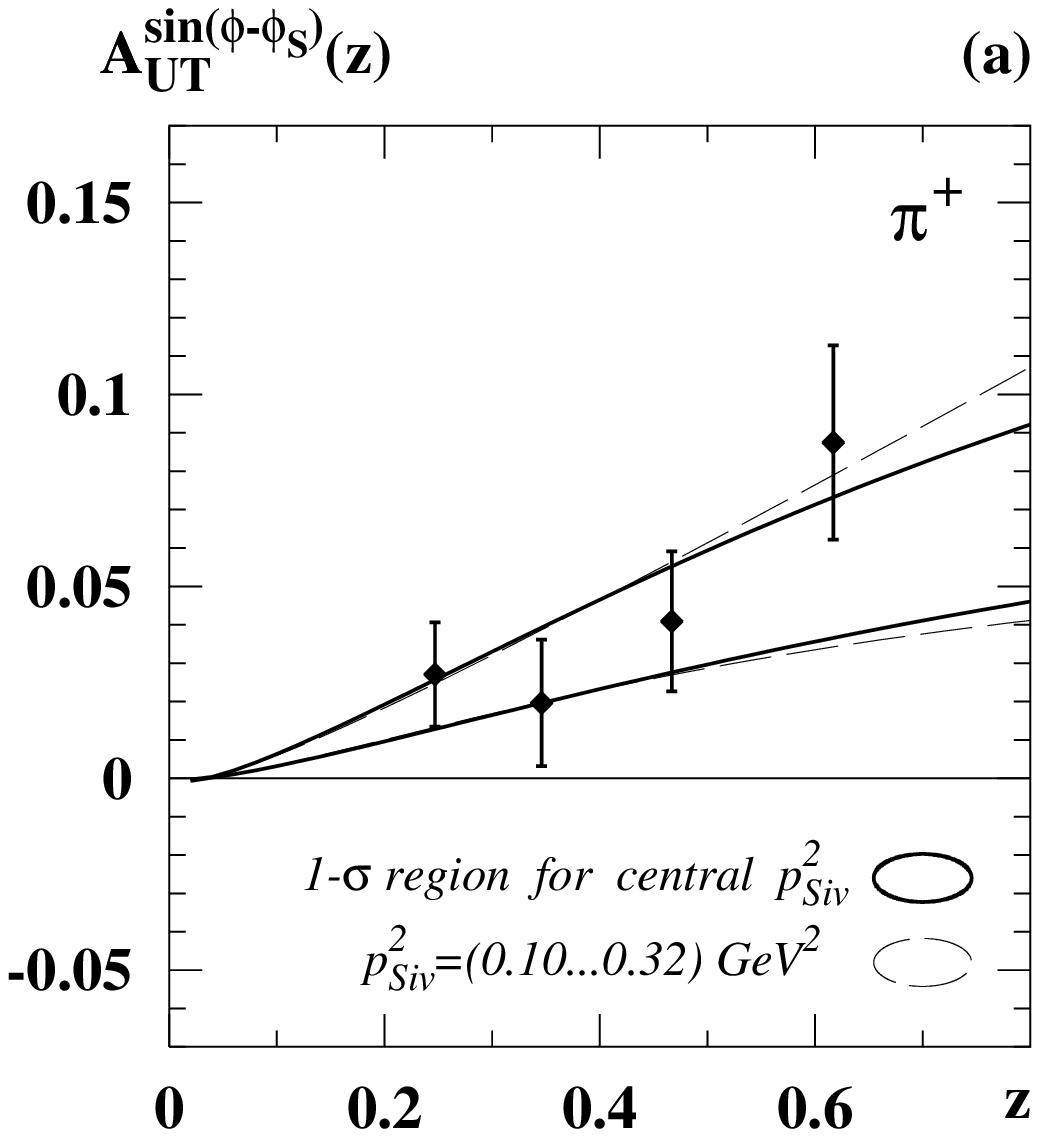}&
\includegraphics[width=2.2in]{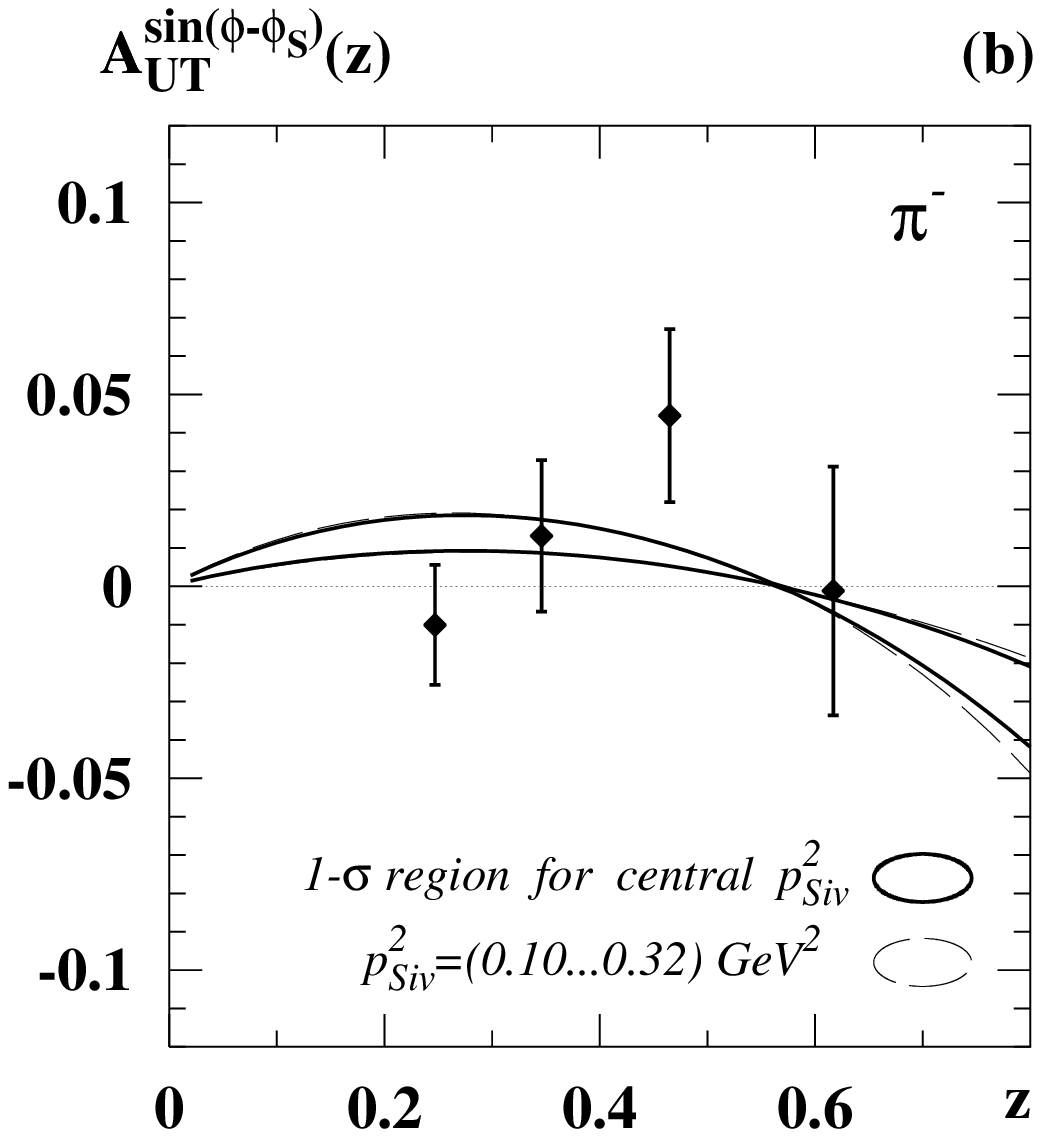}
\end{tabular}
\caption{\footnotesize\sl
    \label{Fig6-AUT-z}
        $A_{UT}^{\sin(\phi_h-\phi_S)}$ as function of $z$.
        The data are from the HERMES experiment \cite{Airapetian:2004tw}.
        The curves show the 1-$\sigma$ variation of our large-$N_c$ 
        constrained fit (\ref{Eq:2nd-fit}) for the Sivers function,
        and the impact of varying the Gaussian width $p^2_{\rm Siv}$
        of the Sivers function in the range (\ref{Eq:pSiv2-range}).
        The z-dependent data were not used for the fit, i.e., the
        comparison serves as a cross check for our results.}
\end{figure}

In our fitting procedure we did not use the HERMES data 
\cite{Airapetian:2004tw} on the $z$-dependence of the Sivers SSA. 
These data could have been used as an additional constraint for the 
integrals of $x\,f_{1T}^{\perp (1) a}(x)$ in the range $0.023 < x < 0.4$,
which corresponds to the cuts in the HERMES experiment. This would have helped
to improve the significance of the fit, considering that only few $x$-data 
points are available.
Instead, let us use these data here as a valuable cross check of our
approach. As the $z$-shape of the SSA is dictated by the unpolarized
fragmentation function $D_1^a(z)$ {\sl and} the $z$-dependence of the 
Gaussian factor $a_{\rm Gauss}$ in Eq.~(\ref{Eq:AUT-SIDIS-Gauss}), this is 
not only a cross check for the extracted Sivers function
(\ref{Eq:2nd-fit}), but it  also tests the Gauss ansatz
(\ref{Eq:Gauss-ansatz}), the consistency of  
the choice of parameters (\ref{Eq:fit-pT2-KT2},~\ref{Eq:pSiv2-range}), and
the large-$N_c$ ansatz (\ref{Eq:large-Nc},~\ref{Eq:ansatz}) itself.
In Fig.~\ref{Fig6-AUT-z} we confront our fit result (\ref{Eq:2nd-fit})
with the $z$-dependent HERMES data on the Sivers SSA \cite{Airapetian:2004tw}.
We observe a satisfactory agreement. Notice that the impact of the poorly 
constrained Gaussian width $p^2_{\rm Siv}$ of the Sivers function 
(\ref{Eq:pSiv2-range}) is marginal.

%
%

The smallness of the $\pi^-$ SSA can be explained as follows. Since 
$f_{1T}^{\perp u}=-f_{1T}^{\perp d}$ in the large-$N_c$ limit,
the asymmetry $A_{UT}^{\sin(\phi_h-\phi_S)}(\pi^-)\propto$ 
$(\frac49 D_1^{\rm unf}-\frac19 D_1^{\rm fav})$,
where $D_1^{\rm fav}=D_1^{d\to\pi^-}=D_1^{\bar u\to\pi^-}=D_1^{u\to\pi^+}$,
etc. ($D_1^{\rm unf}=D_1^{u\to\pi^-}=D_1^{d\to\pi^+}$, etc.),
denotes the so-called favoured (unfavoured) fragmentation function.
At any $z$ we have the inequality $D_1^{\rm unf}(z)<D_1^{\rm fav}(z)$. 
However, due to the weighting by the square of the quark electric charges, 
the effects of the 
smaller ``unfavoured'' and the larger ``favoured'' fragmentation function 
become comparable and tend to cancel each other. As a result the SSA for 
$\pi^-$ appears rather small, becoming zero around $z=0.56$
(for the parameterization in \cite{Kretzer:2001pz} at $Q^2=2.5\,{\rm GeV}^2$).
The tendency for cancellation persists with inclusion of the $1/N_c$
corrections, 
which, however, shift the position of the zero --- see 
Sec.~\ref{Sec-7:qbar+Nc-corr}. 

Results from the COMPASS experiment \cite{Alexakhin:2005iw} yield an equally 
important confirmation for the large $N_c$ picture of the Sivers function.
In the COMPASS experiment a solid polarized $^6$LiD target 
\cite{Ball:2003vb,Goertz:2002vv} was used. Neglecting, to a first 
approximation, nuclear binding effects and using isospin symmetry, 
we observe that for the deuterium Sivers distribution function we have 
$f_{1T}^{u/D}\approx f_{1T}^{u/p}+f_{1T}^{u/n}\approx f_{1T}^u+f_{1T}^d$,
and analogously for $d$, $\bar q$, etc.
Thus, the deuterium target is sensitive to the flavour combination
which is suppressed in the large-$N_c$ limit, see (\ref{Eq:large-Nc-0}), 
and for which our ansatz (\ref{Eq:ansatz}) yields exactly zero.
This is in agreement with the present COMPASS data which shows a Sivers 
effect from deuterium target compatible with zero within error bars 
\cite{Alexakhin:2005iw}.

\section{\boldmath Where are Sivers antiquarks \& $1/N_c$ corrections?}
\label{Sec-7:qbar+Nc-corr}

In the ansatz (\ref{Eq:ansatz}) we neglected the Sivers distributions
for antiquarks and for the strange and heavier quarks. The neglect of
strange and heavier 
quarks is probably a good assumption at the present stage of art.
However, neglecting the Sivers antiquarks need not be such a good
approximation.  This can be seen from the unpolarized distribution
functions, since in the region of $x\lesssim 0.15$ the distribution of unpolarized 
$\bar d$-quarks reaches $25\%$ and more of the unpolarized $d$-quark
distribution.  
In fact, this is precisely the $x$-region where HERMES \cite{Airapetian:2004tw}
observes the most significant Sivers effect (for $\pi^+$) --- see 
Fig.~\ref{Fig5-compare-to-data}.

In order to gain a rough idea of the possible uncertainty introduced
by neglecting the Sivers antiquark distributions, we make two simple
models for the Sivers $\bar{u}$- and $\bar{d}$-distributions, while
keeping the quark distributions at the value given by our fit result
(\ref{Eq:2nd-fit}).  Model I is that Sivers $\bar{q}$-distributions
are just $\pm 25\%$ of the corresponding Sivers quark distributions.
This assumption may be an overestimate in the region of larger $x$.
Therefore, in model II we set the ratio of each Sivers $\bar{q}$- to
Sivers $q$-distribution to be the same as for the unpolarized
distributions.  Thus, we will explore the effects of assuming each of
the following {\sl models} for the $\bar q$-Sivers distributions:
\be
\label{Eq:model-Sivers-qbar} 
    f_{1T}^{\perp(1) \bar q}(x) = \epsilon(x) \; f_{1T}^{\perp(1) q}(x) \,, \;\;\;\;\; 
   \mbox{with}\;\;\;\;\; \epsilon(x) =
    \pm\,\cases{ 0.25 = {\rm const} &model I \,, \cr
               \cr
               \displaystyle
                   \frac{(f_1^{\bar u}+f_1^{\bar d})(x)}
                        {(f_1^u+f_1^d)(x)}
              &model II \,.
             }
\ee 
Note that the particular ansatz for model II ensures compatibility
with the large-$N_c$ limit, 
where $(f_1^u+f_1^d)(x)\sim {\cal O}(N_c^2)\gg(f_1^u-f_1^d)(x)\sim {\cal
  O}(N_c)$.  Thus, our model Sivers antiquarks
satisfy large-$N_c$ relations analogous to
(\ref{Eq:large-Nc-0},~\ref{Eq:large-Nc}).  We also
automatically preserve the sum rule \cite{Burkardt:2003yg} 
(see also \cite{Efremov:2004tp}),
\be
\label{Eq:th-05}
     \sum_{a=g,u,d,\,\dots} \int\!\di x \;f_{1T}^{\perp(1)a}(x) = 0\;.
\ee
In the large-$N_c$ limit the gluon Sivers distribution
is suppressed with respect to the quark one \cite{Efremov:2004tp}.
Our models (\ref{Eq:model-Sivers-qbar})
satisfy the inequality corresponding to
(\ref{Eq:positivity-bound-3}).  Thus, being compatible with all
theoretical constraints we are presently aware of, the models
(\ref{Eq:model-Sivers-qbar}) are well suited for our purposes.

%
\begin{figure}
\begin{tabular}{cc}
\includegraphics[width=2.2in]{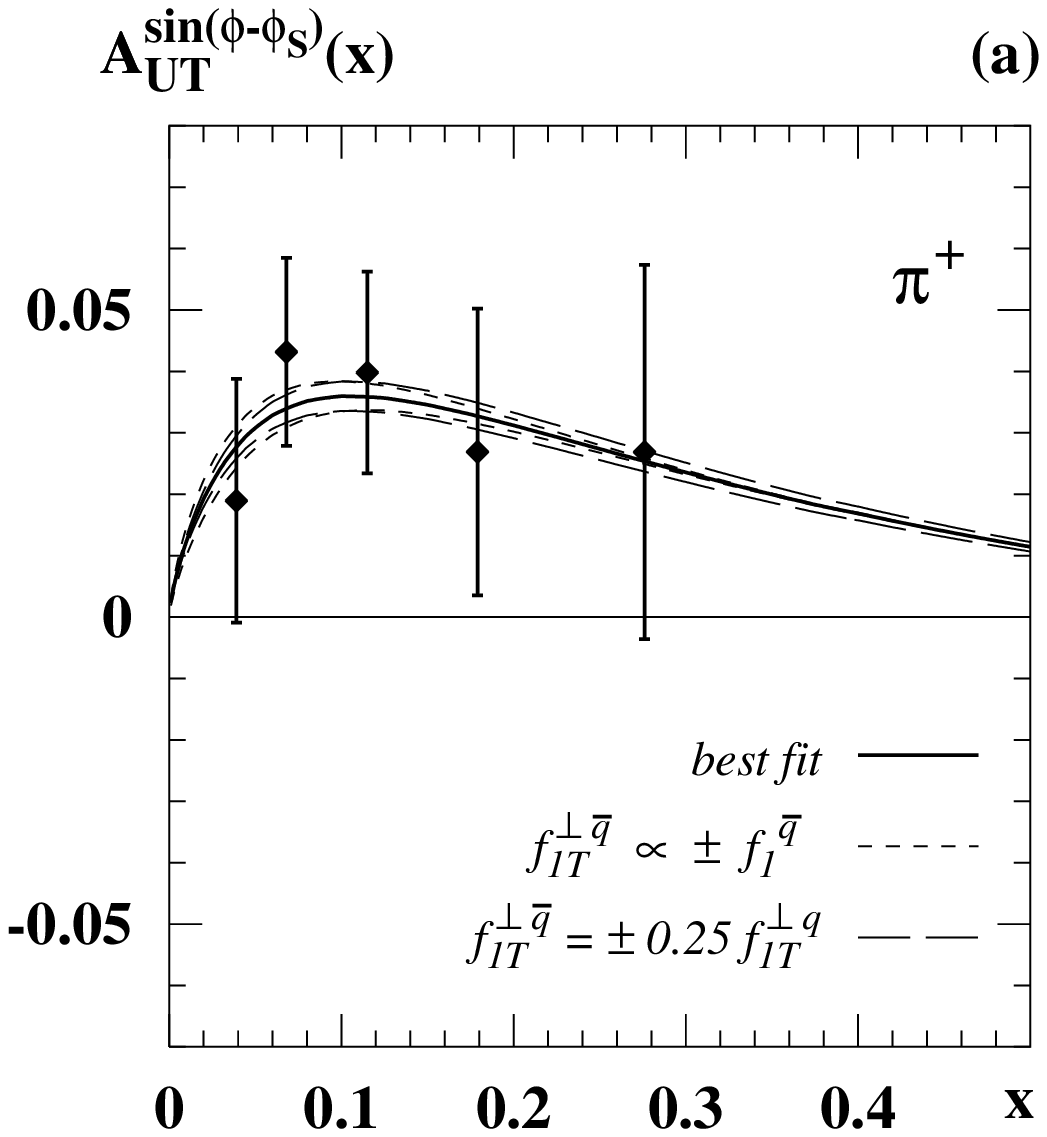}&
\includegraphics[width=2.2in]{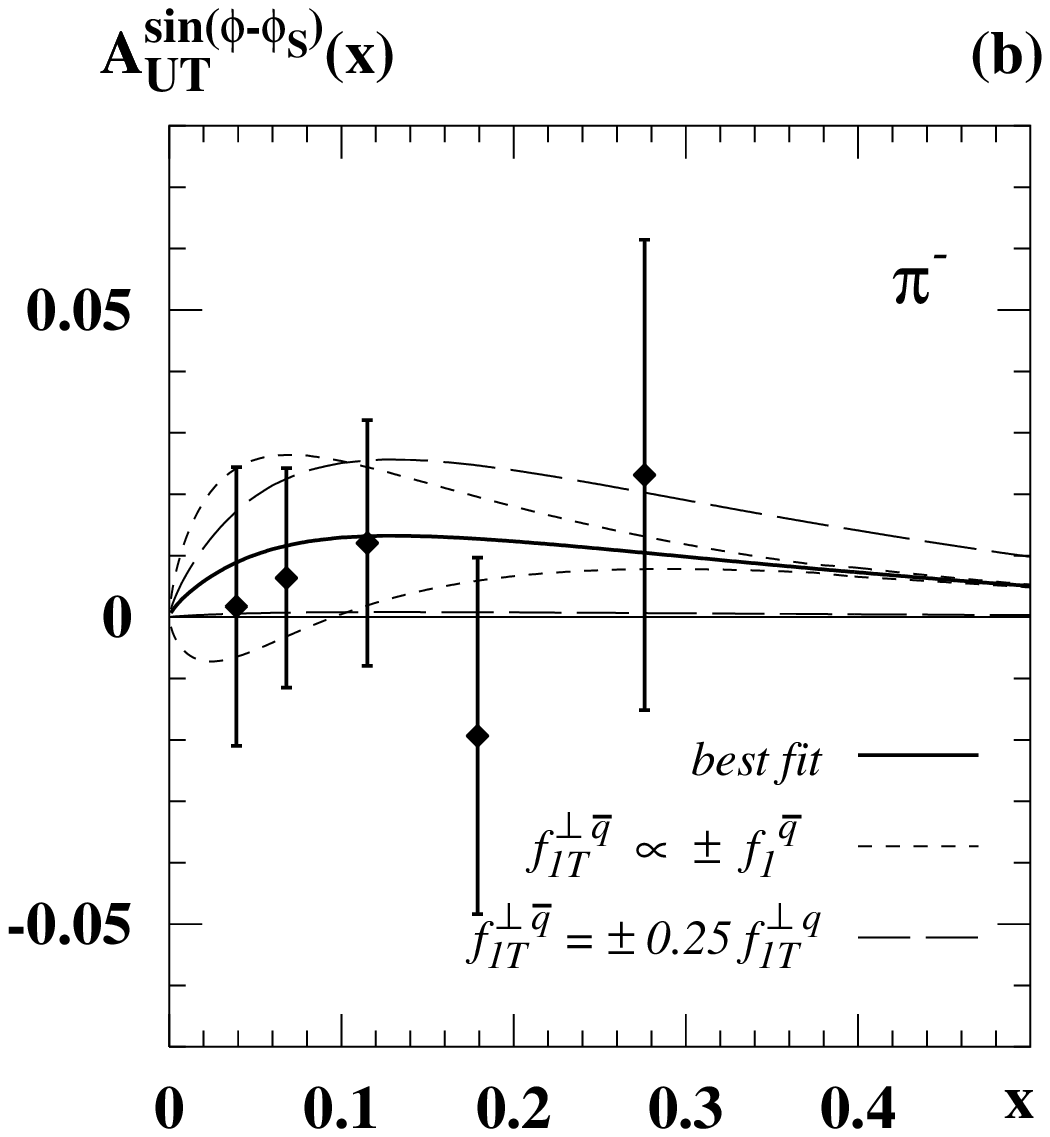} 
\end{tabular}
\caption{\label{Fig7-AUT-Qbar}\footnotesize\sl
        $A_{UT}^{\sin(\phi_h-\phi_S)}$ as a function of $x$.
        The data are from the HERMES experiment \cite{Airapetian:2004tw}.
        The curves show the effect of modifying the result of large-$N_c$ 
        constrained fit (\ref{Eq:2nd-fit}) by using the Sivers
        antiquark distribution functions given in
        Eq.~(\ref{Eq:model-Sivers-qbar}). 
        The figures show that the data \cite{Airapetian:2004tw} 
        have little sensitivity to the Sivers ${\bar q}$-distributions,
        so that the neglect of the antiquark distributions in the fit 
        was justified.}
\end{figure}

In Fig.~\ref{Fig7-AUT-Qbar} we show the modest effect of the Sivers
${\bar q}$-distributions as a function of $x$.  The effect is quite
small for the $\pi^+$ SSA, and well within the experimental errors.
The effect is more pronounced for the $\pi^-$ SSA, but still remains
within the experimental errors.  The effect of Sivers antiquarks on
the $z$-dependence of the Sivers SSA is less visible, and we refrain
from showing analog plots. As here the $x$-dependence is integrated
over, the entire effects amount of altering the overall normalization
of the SSA without qualitative and with only small quantitative
changes to the picture in Fig.~\ref{Fig6-AUT-z}.

Thus, we conclude that even sizeable Sivers antiquark distributions, as
modeled in Eq.~(\ref{Eq:model-Sivers-qbar}), cannot be resolved within the 
error bars of the present data \cite{Airapetian:2004tw}.
This justifies a posteriori the neglect of Sivers ${\bar q}$-distribution 
functions in our fit ansatz (\ref{Eq:ansatz}) here or in 
Ref.~\cite{Efremov:2004tp}. In this way we confirm also the observation 
made in Ref.~\cite{Anselmino:2005nn}.
There an attempt was made to extract $f_{1T}^{\perp a}$ for the separate 
flavours $a=u$, $d$, $\bar{u}$ and $\bar{d}$, and the Sivers-$\bar{q}$ 
distributions were found to be consistent with zero with large
uncertainties.  

%
\begin{figure}
\begin{tabular}{cc}
\includegraphics[width=2.2in]{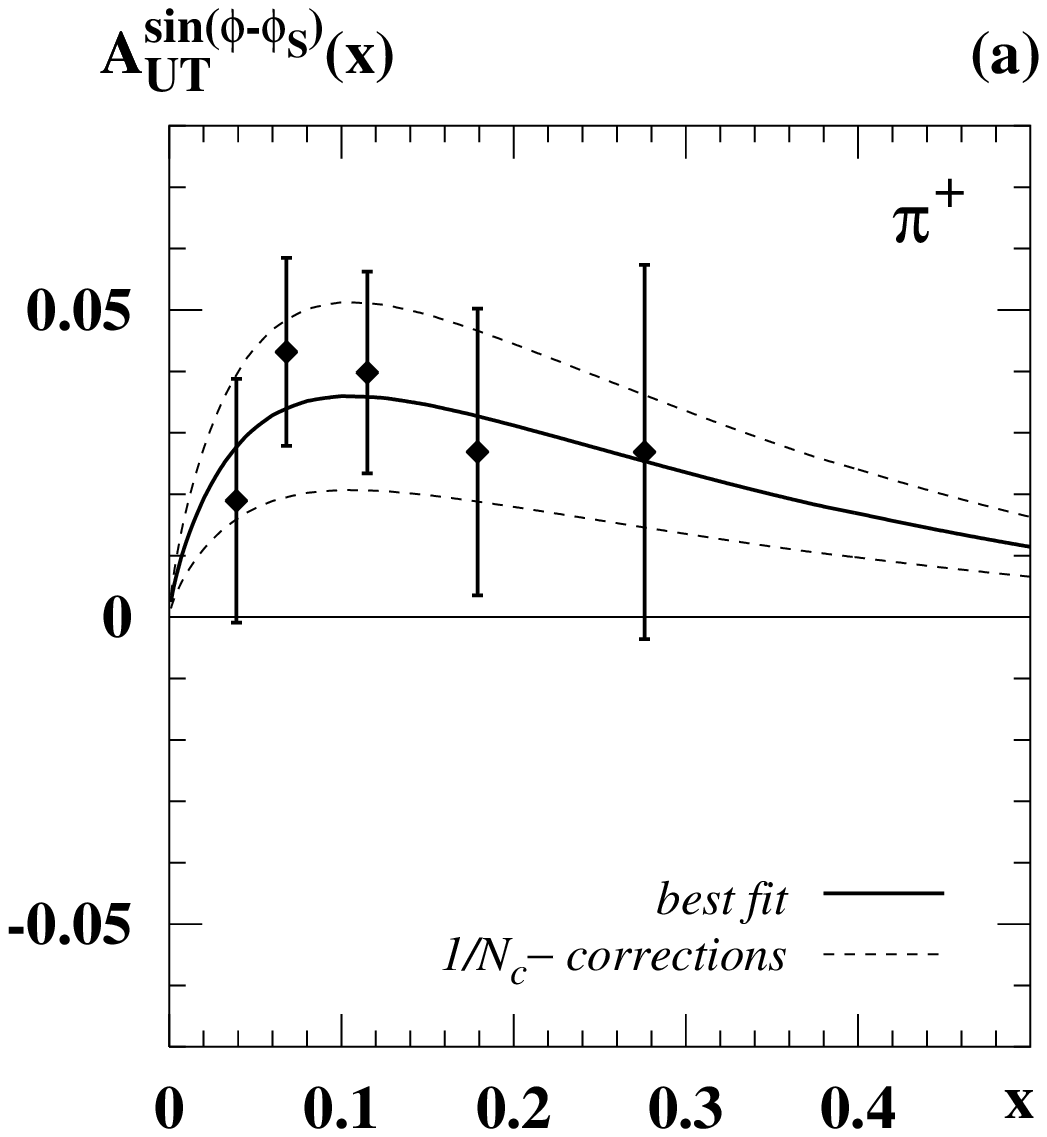}&
\includegraphics[width=2.2in]{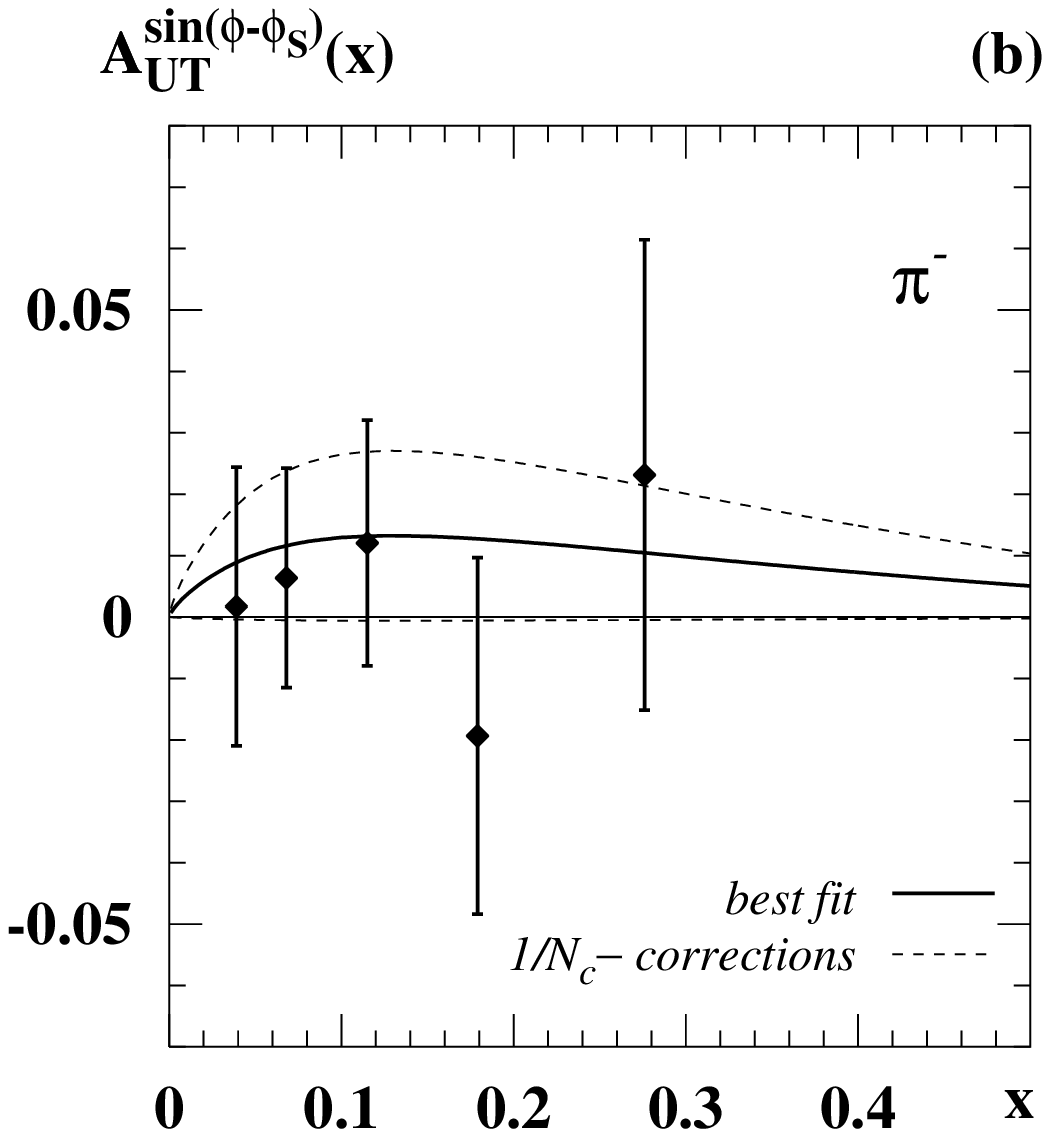}\\
\includegraphics[width=2.2in]{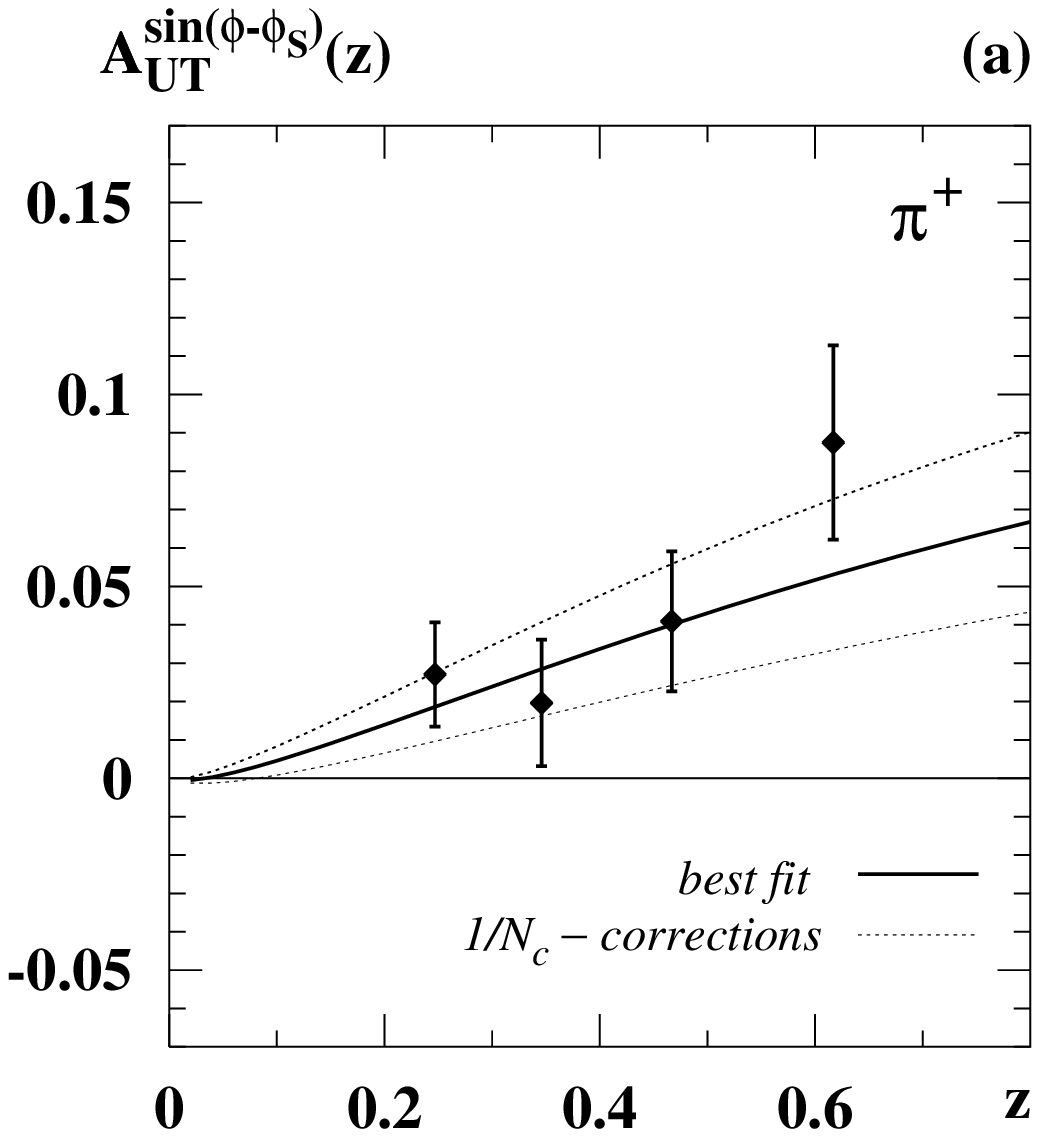}&
\includegraphics[width=2.2in]{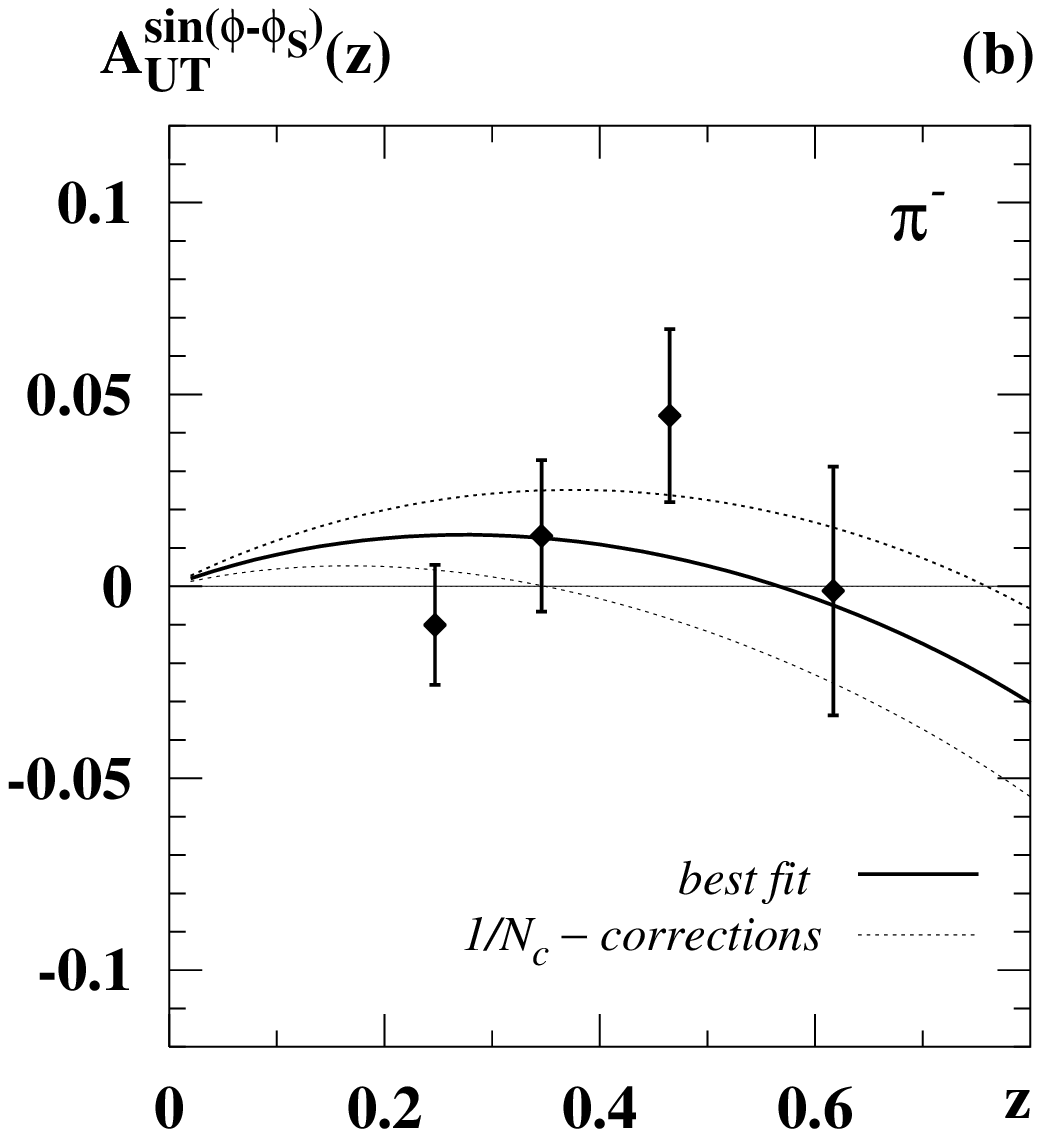} 
\end{tabular}
\caption{\label{Fig8-AUT-Nc-corr}\footnotesize\sl
        {\bf a} and {\bf b}.
        $A_{UT}^{\sin(\phi_h-\phi_S)}$ as function of $x$.
        The data are from the HERMES experiment \cite{Airapetian:2004tw}.
        The curves show our best fit (\ref{Eq:2nd-fit}) for the Sivers 
        function, and the effect of $1/N_c$-corrections as modeled in 
        Eq.~(\ref{Eq:model-Nc-corr}).
        The figures show that the statistical uncertainty of the data 
        \cite{Airapetian:2004tw} is of comparable magnitude as the 
        $1/N_c$-corrections. 
        {\bf c} and {\bf d}. 
        The same as in Figs.~\ref{Fig8-AUT-Nc-corr}a and 
        \ref{Fig8-AUT-Nc-corr}b but for the $z$-dependence
        of the Sivers SSA.}
\end{figure}

\begin{figure}
%
%
\begin{tabular}{cc}
\includegraphics[width=2.2in]{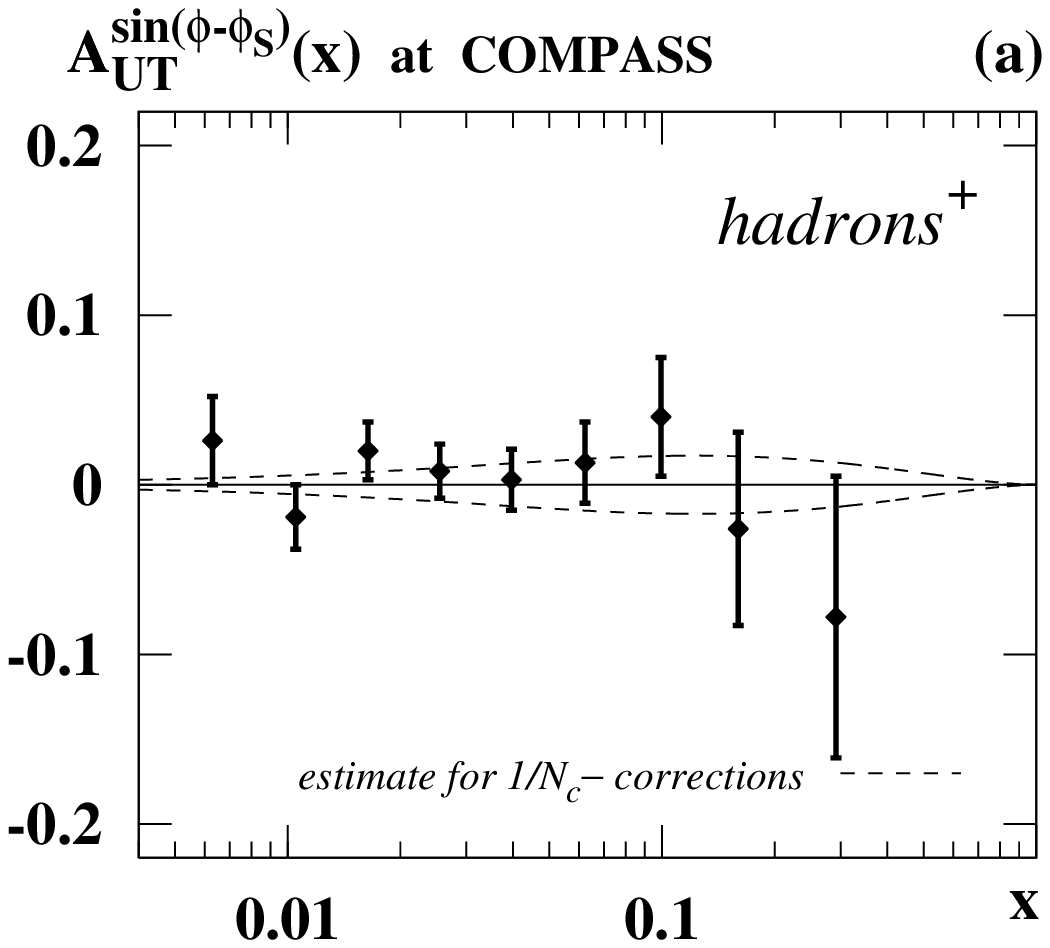}&
\includegraphics[width=2.2in]{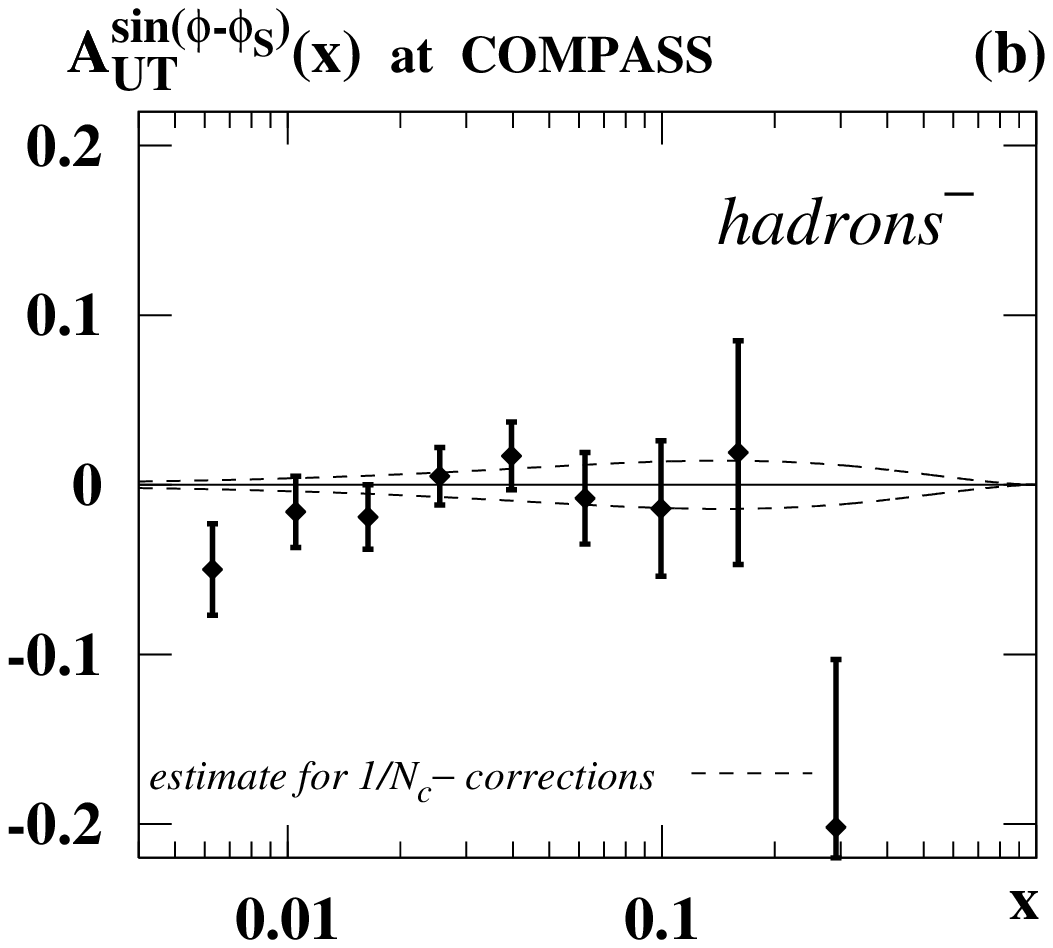} 
\end{tabular}
\caption{\label{Fig9-AUT-Nc-COMPASS}\footnotesize\sl
        The Sivers SSA at COMPASS. The theoretical curves indicate the 
        order of magnitude of the effect according to the rough model
        for $1/N_c$-corrections in Eq.~(\ref{Eq:model-Nc-corr}).
        Thus, taking the large-$N_c$ counting rule (\ref{Eq:large-Nc-0}) 
        into account we conclude that the HERMES \cite{Alexakhin:2005iw} and 
        COMPASS \cite{Alexakhin:2005iw} data are compatible with each other.}
\end{figure}

The difficulty to access antiquark distribution functions
in SIDIS data is more general than the example encountered here. For
example, in 
longitudinal double spin asymmetries $A_{LL}\propto\sum_ae_a^2g_1^a(x)D_1^a(z)$
it is presently \cite{Airapetian:2004zf} not possible to resolve a flavour 
asymmetry in the helicity sea as sizeable as predicted in the chiral 
quark-soliton model \cite{Diakonov:1996sr}, namely larger
\cite{Dressler:1999zg} than the known 
flavour asymmetry in the unpolarized sea 
$(g_1^{\bar u}-g_1^{\bar d})(x) > |(f_1^{\bar u}-f_1^{\bar d})(x)|$.
As in the case of the much better known distribution functions $f_1^a(x)$ and 
$g_1^a(x)$ (see \cite{Dressler:1999zv} for the specific example of 
helicity sea flavour asymmetry from the chiral quark-soliton model), 
it is necessary to study the Drell-Yan process in order to learn 
more about Sivers antiquark distribution functions.

Next let us address the $1/N_c$-corrections. In our ansatz
(\ref{Eq:ansatz}) we took literally the prediction from the large-$N_c$
limit \cite{Pobylitsa:2003ty} in Eq.~(\ref{Eq:large-Nc}), disregarding
corrections which are generically of ${\cal O}(1/N_c)\approx 30\%$.
In order to have an idea of the effect of these corrections,
let us assume that the flavour singlet Sivers distribution is not exactly zero
but suppressed by exactly a factor of $1/N_c$ with respect to the
flavour non-singlet 
combination according to Eq.~(\ref{Eq:large-Nc-0}). That is,
\be\label{Eq:model-Nc-corr}
        \Big |(f_{1T}^{\perp(1)u}+f_{1T}^{\perp(1)d})(x) \Big | \stackrel{!}{=} 
        \pm \;\frac{1}{N_c} (f_{1T}^{\perp(1)u}-f_{1T}^{\perp(1)d})(x) \,,
\ee
where we use $f_{1T}^{\perp(1)u}(x)$, $f_{1T}^{\perp(1)d}(x)$ from 
Eq.~(\ref{Eq:2nd-fit}) and set $N_c=3$. Of course, it would be naive 
to expect that the different flavour combinations behave precisely
as in Eq.~(\ref{Eq:model-Nc-corr}). However, the scenario
in Eq.~(\ref{Eq:model-Nc-corr}) is by construction 
well suited to indicate the order of magnitude of the effect. In fact,
when considered from such a qualitative point of view, large-$N_c$
relations for parton distributions are observed to be well satisfied 
in nature \cite{Efremov:2000ar}.

Note that (\ref{Eq:model-Nc-corr}) is compatible with positivity
(\ref{Eq:positivity-bound-3}). However, in order to comply with the
sum rule (\ref{Eq:th-05}), we must
introduce a gluon Sivers distribution function equal in magnitude but 
of opposite sign to the quark flavour singlet Sivers distribution.

Fig.~\ref{Fig8-AUT-Nc-corr} shows the effect of $1/N_c$-corrections as 
modeled in Eq.~(\ref{Eq:model-Nc-corr}).
The positive (negative) sign in Eq.~(\ref{Eq:model-Nc-corr}) corresponds 
to the upper (lower) curves in Figs.~\ref{Fig8-AUT-Nc-corr}a and 
\ref{Fig8-AUT-Nc-corr}b. 
What we obtained in this way is an ``error band'' of much the same size as 
the 1-$\sigma$ uncertainty of our fit in Fig.~\ref{Fig5-compare-to-data}.
A look at the $z$-dependence of the Sivers SSA fully confirms these
findings --- see Figs.~\ref{Fig8-AUT-Nc-corr}c and \ref{Fig8-AUT-Nc-corr}d.
        When the HERMES $z$-cuts are used,
        $ \la a_{\rm Gauss}D_1^{\rm fav}\ra\approx $
        $2\la a_{\rm Gauss}D_1^{\rm unf}\ra$ to a good accuracy with 
	the parameterization \cite{Kretzer:2001pz}. If this were exactly the case, then
	for the minus sign in Eq.~(\ref{Eq:model-Nc-corr}) the $\pi^-$ 
        Sivers SSA would become exactly zero, as shown in
        Fig.~\ref{Fig8-AUT-Nc-corr}b.  
More generally, we see that the vanishing of the $\pi^-$ asymmetry at
some value of $z$ is a feature which is robust against
$1/N_c$-corrections (and  antiquark effects). The precise position of
this zero, however, is very 
sensitive to $1/N_c$-corrections --- Fig.~\ref{Fig8-AUT-Nc-corr}d.

%
        \begin{wrapfigure}{RD!}{6cm}
        \centering
        \includegraphics[width=2.0in]{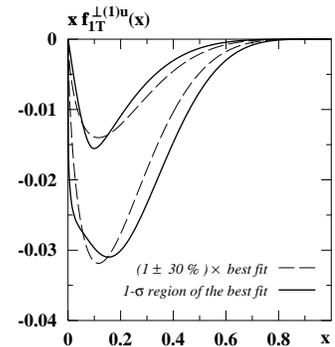}
        \caption{\footnotesize\sl
        \label{FigXX}
	The 1-$\sigma$ region of our best fit for $f_{1T}^{\perp(1)u}(x)$ 
        in comparison to a variation of $\pm30\%$ of our best fit 
        (\ref{Eq:2nd-fit}), which roughly corresponds to the generic 
	size of $1/N_c$ corrections.}
        \end{wrapfigure}
%
On a deuteron target, the leading $1/N_c$ prediction gives zero for
the SSA, so that the $1/N_c$ corrections are all that remain.
Assuming for simplicity that the positive and negative hadrons identified at 
COMPASS are mainly pions, we obtain in our rough model 
(\ref{Eq:model-Nc-corr}) for $1/N_c$-corrections the results shown 
in Fig.~\ref{Fig9-AUT-Nc-COMPASS}. The positive asymmetries are due
to choosing the positive sign in Eq.~(\ref{Eq:model-Nc-corr}).
Clearly, we see that the COMPASS data \cite{Alexakhin:2005iw} are compatible 
with the large-$N_c$ corrections being of a magnitude compatible with
out model.

Thus, the reason why our large-$N_c$ approach works here, is due to 
the fact that current precision of the first experimental data 
\cite{Airapetian:2004tw,Alexakhin:2005iw} is comparable to the theoretical 
accuracy of the large-$N_c$ relation (\ref{Eq:large-Nc}). 
This is illustrated in a different way by Fig.~\ref{FigXX} which shows that
a variation of the best fit (\ref{Eq:2nd-fit}) of $\pm30\%$ corresponding 
to the generic size of $1/N_c$ corrections is of similar size and shape
as the 1-$\sigma$ region of the fit.
In other words, 
$1/N_c$ corrections (and antiquark effects) cannot be resolved within the 
error bars of the data \cite{Airapetian:2004tw,Alexakhin:2005iw}. In future, 
with increasing precision of the data, it will certainly be necessary to 
refine the fit ansatz (\ref{Eq:ansatz}) to include $1/N_c$ corrections and 
antiquarks. In fact we have found that our fit (\ref{Eq:2nd-fit}) to the 
final HERMES data \cite{Airapetian:2004tw} is also compatible with the 
most recent and substantially more precise {\sl preliminary} HERMES
data given in \cite{Diefenthaler:2005gx}.

\section{Further tests of the Gaussian ansatz in SIDIS}

We have seen that the Gaussian model (\ref{Eq:Gauss-ansatz}) for the 
distribution of the transverse parton momenta provides a satisfactory 
effective description of the HERMES data on the average transverse momentum 
of pions produced in SIDIS \cite{Airapetian:2002mf}. 
In order to increase our faith into the applicability of the Gaussian model and
--- what is equally important --- to find out its limitations, it is necessary 
to make further tests using, e.g., the available HERMES data both on 
polarized and unpolarized SIDIS. 
Although the true prediction in QCD, or any other field theory, is
that there is approximately a power-law fall off at large transverse
momentum, it can well be that a Gaussian dependence can be a useful
approximation
for transverse momenta that are low with respect to the relevant hard
scale.  
This, in fact, corresponds to the situation in the HERMES experiment 
\cite{Airapetian:2004tw} where 
$\la P_{h\perp}\ra\sim 0.4\,{\rm GeV}\ll\sqrt{\la Q^2\ra}\sim 1.5\,{\rm GeV}$.

In this context it would be interesting to study the average transverse 
momentum square $\la P_{h\perp}^2(z)\ra$ of the produced hadrons given by 
(actually this relation is of more general character \cite{Bacchetta:2002tk}
and manifestly valid in the Gauss ansatz)
\be\label{Eq:Phperp2-av}
        \la P_{h\perp}^2(z)\ra = z^2 p^2_{\rm unp}+K^2_{\!D_1}\:.
\ee
In particular, $\la P_{h\perp}(z)\ra$ and $\la P_{h\perp}^2(z)\ra$ are
related to each other in the Gaussian model by
\be
        \la P_{h\perp}(z)\ra^2 \stackrel{\rm Gauss}{=} 
        \frac{\pi}{4}\;\la P_{h\perp}^2(z)\ra\;.
\ee
To test the assumption of flavour independence, it would
be useful to study $\la P_{h\perp}(z)\ra$ and $\la P_{h\perp}^2(z)\ra$
separately for $\pi^+$, $\pi^0$, $\pi^-$ or the kaons. Finally, one could
study the average transverse momentum of different hadrons averaged over 
$z$ but as function of the respective $x$-bin, in order to test the 
assumption of an $x$- and flavour independent Gaussian width of the 
unpolarized distribution function.

Given the unsatisfactory situation with the poorly constrained Gaussian
width of the Sivers function, it is of importance to further constrain this 
parameter by means 
of data --- in particular on the Sivers SSA as function of $P_{h\perp}$.
Appropriate HERMES data for this as well as the following suggestion are 
in principle available~\cite{Airapetian:2004tw,HERMES-new}.

Let us define the following ``Sivers-mean-transverse-momentum'':
\be
        \la P_{h\perp}(z)\ra_{\rm Siv} = \frac{
         \sum_iP_{h\perp,i}\;\sin(\phi_i-\phi_{S,i})
         \left[ N^\uparrow(\phi_i;\phi_{S,i})-N^\downarrow(\phi_i;\phi_{S,i}+\pi) \right] }
        {\sum_i\sin(\phi_i-\phi_{S,i})
         \left[ N^\uparrow(\phi_i;\phi_{S,i})-N^\downarrow(\phi_i;\phi_{S,i}+\pi) \right] }
        \stackrel{\rm Gauss}{=}
        \frac{2}{\sqrt{\pi}}\;\sqrt{z^2 p^2_{\rm Siv}+K^2_{\! D_1}}\;.
\ee

\noindent 
        \begin{wrapfigure}{RD!}{6cm}
       \vspace{0.7cm}
        \centering
        \includegraphics[width=2.0in]{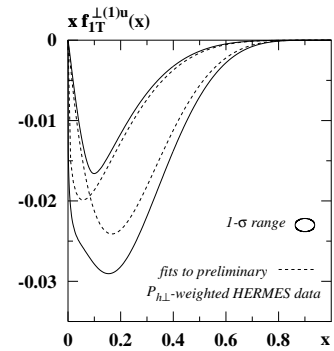}
        \caption{\footnotesize\sl
        \label{Fig10-compare-Phperp-weighted-data}
        The fits from Ref.~\cite{Efremov:2004tp} of $f_{1T}^{\perp(1)a}(x)$ 
        to the $P_{h\perp}$-weighted {\sl preliminary} HERMES data 
        \cite{HERMES-new} lie within the 1-$\sigma$ region of the fit 
        obtained here to the {\sl final} (non-$P_{h\perp}$-weighted) HERMES
        data \cite{Airapetian:2004tw}.}
        \end{wrapfigure}

One can read off the result in the Gaussian model by comparing 
Eqs.~(\ref{Eq:04}) and (\ref{Eq:AUT-SIDIS-Gauss}). Note that 
$\la P_{h\perp}(z)\ra_{\rm Siv}$ is of the same structure as its 
``unpolarized analog'' $\la P_{h\perp}(z)\ra$ in Eq.~(\ref{Eq:mean-momentum})
up to the coefficient $2/\sqrt{\pi}$. 
From a simultaneous analysis of the Sivers SSA weighted with and without
a power of transverse hadron momentum $\la P_{h\perp}(z)\ra_{\rm Siv}$ could 
be determined with a relative accuracy comparable to that of the Sivers SSA.

An indirect but important test of the Gaussian ansatz for the Sivers 
function can be made on the basis of the analyses reported here and in 
\cite{Efremov:2004tp}. 
In Ref.~\cite{Efremov:2004tp} the {\sl preliminary} HERMES data 
\cite{HERMES-new} on the transverse-momentum-weighted Sivers SSA (\ref{Eq:04}) 
were used in order to directly extract $f_{1T}^{\perp(1)a}(x)$ without 
resorting to any model for the transverse parton momenta.
Here we have used the {\sl final} HERMES data \cite{Airapetian:2004tw} to 
extract $f_{1T}^{\perp(1)a}(x)$ {\sl under the assumption} of the Gaussian 
model (\ref{Eq:Gauss-ansatz}). Thus, if the Gaussian model works, then the 
two extractions must yield the same result. 

In \cite{Efremov:2004tp} the same set of assumptions (large-$N_c$ for $q$,
neglect of $\bar q$, etc.) was used as here.  However, instead of 
determining the 1-$\sigma$ region, two different ans{\"a}tze were explored: 
One ansatz is as in (\ref{Eq:ansatz}), and another is as in
(\ref{Eq:ansatz}) 
but with fixed $b=1$. In Fig.~\ref{Fig10-compare-Phperp-weighted-data} we see 
that these fits from Ref.~\cite{Efremov:2004tp} of $f_{1T}^{\perp(1)a}(x)$ to 
the $P_{h\perp}$-weighted {\sl preliminary} HERMES data \cite{HERMES-new} 
are compatible within 1-$\sigma$ with the fit to the {\sl final} HERMES data 
on the Sivers SSA weighted without a power of $P_{h\perp}$ 
\cite{Airapetian:2004tw}. 

This observation {\sl indicates} that the Gaussian ansatz for the Sivers 
function is compatible with the HERMES data within the statistical accuracy 
of the data \cite{Airapetian:2004tw,HERMES-new}.
Given the {\sl preliminary} status of the data \cite{HERMES-new} it is not 
possible to draw a stronger conclusion from this observation at the present 
stage. In fact, the HERMES Collaboration does not recommend \cite{Delia} 
the use of the preliminary data \cite{HERMES-new} since they are not corrected 
for acceptance effects specific to the $P_{h\perp}$-weighting, i.e., absent 
in the non-$P_{h\perp}$-weighted SSA \cite{Airapetian:2004tw}.
However, our comparison in Fig.~\ref{Fig10-compare-Phperp-weighted-data}
indicates that the systematic error due to these effects is less dominant than
the statistical uncertainty of the data \cite{Airapetian:2004tw,HERMES-new}.

\begin{figure}[h]
\begin{tabular}{cc}
\includegraphics[width=2.2in]{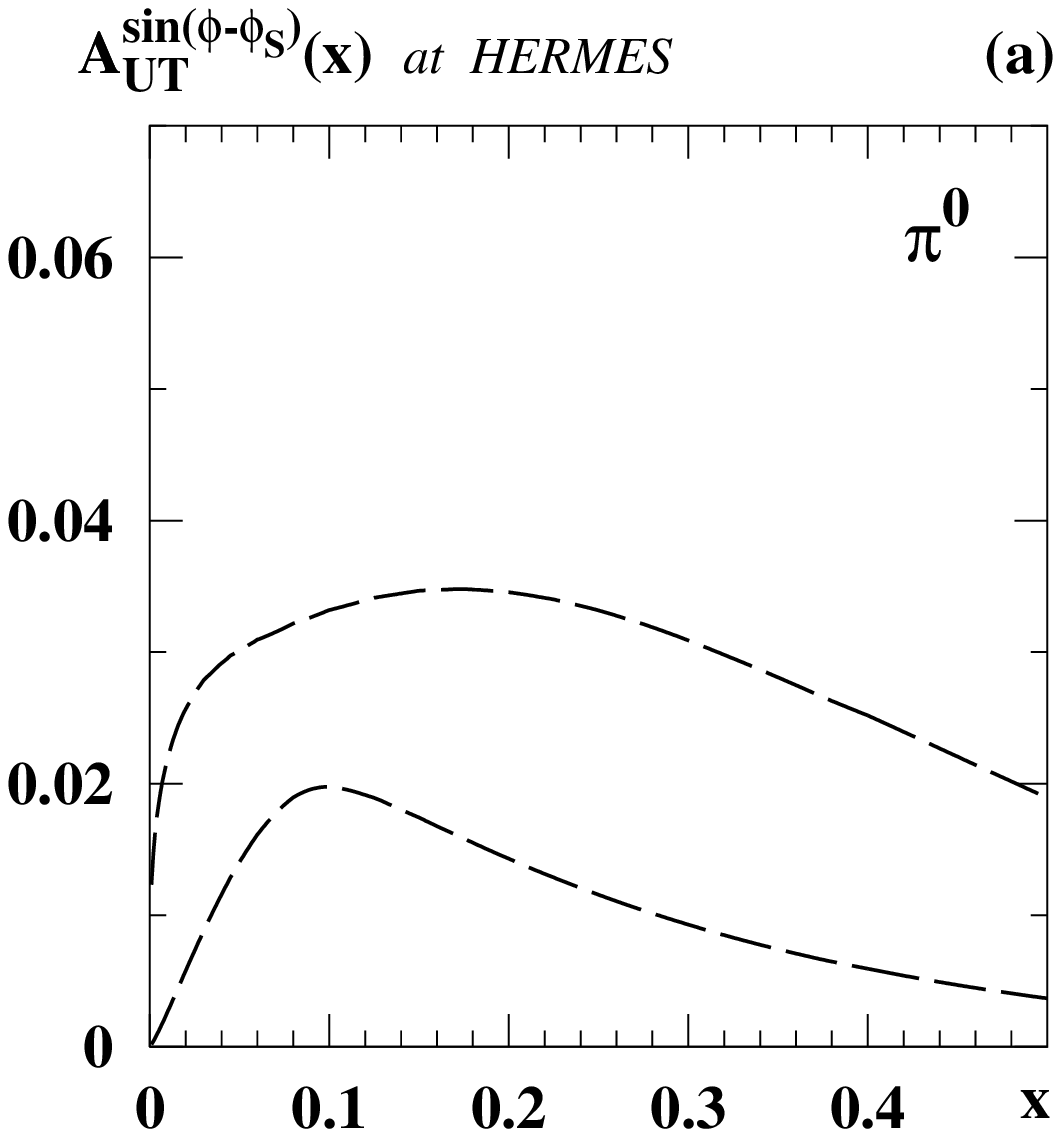}&
\includegraphics[width=2.2in]{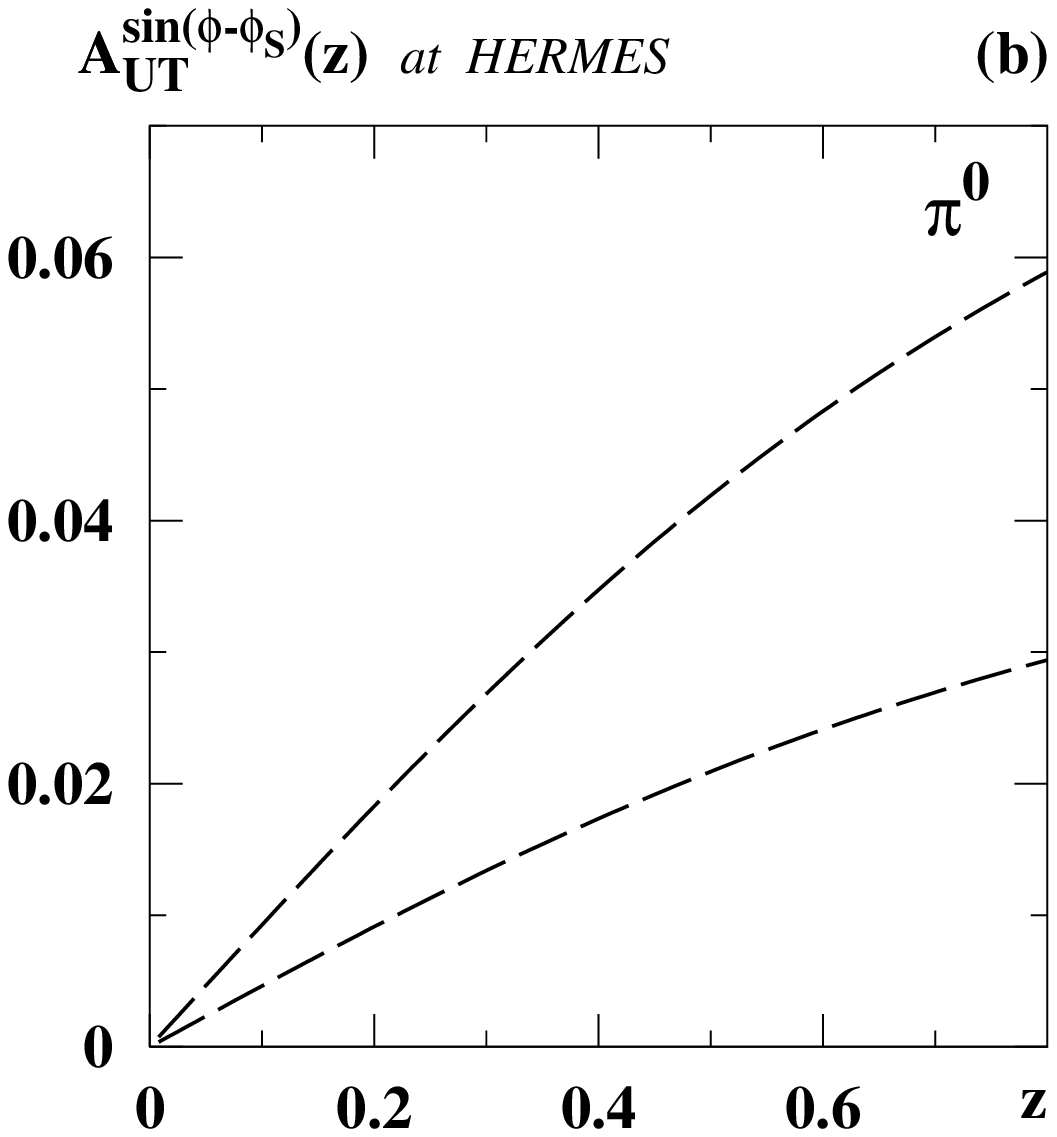} 
\end{tabular}
\caption{\label{Fig11-AUT-HERMES-Pi0}\footnotesize\sl
        The Sivers SSA for neutral pions at HERMES as function of $x$ and 
        $z$, respectively, as predicted on the basis of the fit result 
        (\ref{Eq:2nd-fit}). The error band arises from the 1-$\sigma$
        uncertainty of the fit (\ref{Eq:2nd-fit}).}
\end{figure}

Another important test for the Gaussian model and for the fit result 
(\ref{Eq:2nd-fit}) could use the HERMES data (and possible future data from
COMPASS and Jlab) on the $\pi^0$ Sivers SSA. 
Preliminary data on the $\pi^0$ SSA with large statistical uncertainties were 
shown in \cite{HERMES-new}. Taking into account the 1-$\sigma$ uncertainty of 
the fit (\ref{Eq:2nd-fit}) we obtain the results shown in 
Fig.~\ref{Fig11-AUT-HERMES-Pi0}.
Worthwhile commenting is the $z$-dependence of the $\pi^0$-asymmetry. Since 
the unpolarized fragmentation functions of light quarks and antiquarks into 
$\pi^0$ are the same (and since we neglect the effects of strange and heavier 
quarks), $D_1^a(z)$ completely cancel out from the SSA, so that
$A_{UT}^{\sin(\phi-\phi_S)}(z,\pi^0) \propto a_{\rm Gauss}(z)$. 
Thus the $z$-shape of the neutral pion SSA is entirely predicted by the 
Gaussian model. 
(This remains true even if one considers effects of $P_{h\perp}$ cuts.
The effect of the small resolution cut applied at HERMES or COMPASS
\cite{Airapetian:2004tw,Alexakhin:2005iw} is negligible in 
Fig.~\ref{Fig11-AUT-HERMES-Pi0}, see also Sec.~\ref{Sec-5:Extraction}.)
A precise measurement of the $z$-dependence of the $\pi^0$ 
Sivers effect could therefore also help to test the Gaussian Ansatz.
Alternatively, one may combine appropriately $\pi^+$ and $\pi^-$ data and 
use isospin symmetry to arrive at the same information. This may yield results 
with better statistical accuracy in the HERMES experiment \cite{Andy-remark}.

%
%

\section{Conclusions}

In this work we have extracted the transverse moment $f_{1T}^{\perp(1)a}$
of the Sivers function from SIDIS HERMES data \cite{Airapetian:2004tw}
using a Gaussian model for the distribution of parton transverse momenta
and employing predictions from the large-$N_c$ limit \cite{Pobylitsa:2003ty} 
as an additional constraint.  We have shown that the Gaussian model 
provides a reasonable description of HERMES data on the transverse momentum 
distribution of the hadrons produced in unpolarized SIDIS. We constrained the 
free parameters of the Gaussian model consistently by HERMES data, which
however, does leave the Gaussian width of the Sivers distribution poorly
constrained. Nevertheless, the data \cite{Airapetian:2004tw} well constrain a 
fit of the {\sl transverse moment} of the Sivers function. 

We have also shown that the HERMES and COMPASS data 
\cite{Airapetian:2004tw,Alexakhin:2005iw} are compatible with each other and
with predictions from the large-$N_c$ limit of QCD \cite{Pobylitsa:2003ty} 
within their statistical accuracy. 
We checked explictly that the effects of the simplifications we made
are either negligibly small, as for example the neglect of the
experimental resolution cuts on the transverse hadron momenta, or well
within the statistical accuracy of the data, as the usage of large-$N_c$ 
constraints or the neglect of Sivers antiquark distributions.
We provided cross and consistency checks for the fit result by studying the 
$z$-dependence of the HERMES data on the Sivers SSA \cite{Airapetian:2004tw},
and made suggestions how to further test the applicability of the Gaussian
model in SIDIS.

The main differences of our approach as compared to the similar works 
\cite{Anselmino:2005nn,Anselmino:2005ea,Vogelsang:2005cs} are
the use of the large-$N_c$ constraints, 
and the choice of a different model for transverse parton momenta 
and/or the way we fixed the respective parameters.
Our fit is in qualitative agreement with extractions of the Sivers function
\cite{Anselmino:2005nn,Anselmino:2005ea,Vogelsang:2005cs} from the same 
\cite{Airapetian:2004tw} and from the more recent and more precise 
(but {\sl preliminary}) HERMES data \cite{Diefenthaler:2005gx},
see Ref.~\cite{Anselmino:2005an} for a detailed comparison.

Of particular interest are studies of the Sivers effect in the Drell-Yan (DY) 
process, because the Sivers function (and other 
``time-reversal--odd'' distributions) are expected to obey an unusual 
universality property, namely to appear with opposite signs in SIDIS and in 
DY \cite{Collins:2002kn}. 
The experimental check of this prediction is a crucial test for the 
understanding of the Sivers effect within QCD.
On the basis of an analysis of the {\sl preliminary} HERMES data 
\cite{HERMES-new} it was shown \cite{Efremov:2004tp} that this change of sign 
of the Sivers function could be checked in $p^\uparrow\pi^-\to l^+l^-X$ in the
planned hadron-beam-mode of the COMPASS experiment \cite{COMPASS-proposal}, 
and in $p^\uparrow\bar{p}\to l^+l^- X$ in the proposed PAX experiment 
\cite{PAX} (whose primary goal is to access the transversity distribution 
function in $p^\uparrow\bar{p}^\uparrow\to l^+l^- X$ \cite{PAX-estimates}).
Our study of the {\sl final} HERMES data \cite{Airapetian:2004tw}
presented here confirms the analysis of Ref.~\cite{Efremov:2004tp} 
solidifying the conclusions made there --- as do the estimates reported in 
Refs.~\cite{Anselmino:2005ea,Vogelsang:2005cs} obtained from analyses of 
the most recent {\sl preliminary} HERMES data \cite{Diefenthaler:2005gx}.
Estimates for COMPASS, PAX and RHIC on the basis of the results
obtained here will be presented elsewhere~\cite{in-progress}.

\paragraph*{Acknowledgments.}
We thank Delia Hasch, Gunar Schnell and Alexei Prokudin for valuable 
discussions.
We also thank Andy Miller for pointing out that the {\sl published} data
\cite{Airapetian:2004tw} are not {\sl final} -- contrary to the way we refer
to them here -- since the HERMES transverse target polarization run is not 
finished and more data are expected \cite{Diefenthaler:2005gx}.

The work is partially supported by BMBF and DFG of Germany, the
COSY-J{\"u}lich project, the Transregio Bonn-Bochum-Giessen, and is
part of the European Integrated Infrastructure Initiative Hadron
Physics project under contract number RII3-CT-2004-506078. 
A.E. is supported by grants RFBR 03-02-16816 and DFG-RFBR 03-02-04022.
J.C.C. is supported in part by the U.S. D.O.E., and by a Mercator
Professorship of DFG.

\end{document}